\documentclass[preprint,12pt]{elsarticle}

\usepackage{amsmath}
\usepackage{amssymb}
\usepackage{graphicx}
\usepackage{braket}
\usepackage{natbib}
\usepackage{url}
\usepackage{subfig}
\usepackage{float}
\usepackage{bm}
\usepackage{here}
\usepackage{hyperref}
\usepackage{titlesec}
\titleformat{\subsubsection}[runin]
  {\normalfont\normalsize\bfseries}{\thesubsubsection}{1em}{}
\usepackage{booktabs} % For formal tables
\usepackage{siunitx}  % For aligning numbers by decimal point
\usepackage{multicol}  
\usepackage{longtable}
\usepackage{booktabs}  % For nicer horizontal lines
\makeatletter
\def\ps@pprintTitle{%
  \let\@oddhead\@empty
  \let\@evenhead\@empty
  \def\@oddfoot{\reset@font\hfil\thepage\hfil}
  \let\@evenfoot\@oddfoot
}
\makeatother
\begin{document}

\begin{frontmatter}

\title{Neural Quantum States in Variational Monte Carlo Method: A Brief Summary}
\author[inst1]{Yuntai Song\corref{cor1}}
\ead{yuntais2@illinois.edu}

\address[inst1]{Department of Physics, University of Illinois Urbana-Champaign, 1110 West Green Street, Urbana, IL 61801, USA}

\cortext[cor1]{Corresponding author: Yuntai Song}

\begin{abstract}
Variational Monte Carlo (VMC) methods have found extensive applications in quantum many-body physics and quantum chemistry since the mid-20th century, particularly demonstrating their unique advantages in studying strongly correlated electron systems. The VMC method is based on the variational principle, estimating the ground state energy of quantum systems by optimizing a parameterized wave function. The development and improvement of this method involve several important research stages, each marked by significant contributions.

The earliest research on VMC can be traced back to W.L. McMillan's work in 1965, studying the ground state of liquid helium-4 \cite{PhysRev.138.A442}. His pioneering study not only demonstrated the potential of VMC methods in physics but also provided a powerful tool for subsequent researchers to explore more complex quantum systems.

In the 1980s, VMC methods began to be widely applied in the field of quantum chemistry, especially in simulating complex molecular systems. The work of James B. Anderson in 1975 showcased the application of VMC in chemical physics, particularly in his simulations of the H\(_3^+\) molecule \cite{anderson_randomwalk_1975}.

In 2001, Sandro Sorella proposed the stochastic reconfiguration (SR) method \cite{PhysRevB.64.024512}, a technique for optimizing wave function parameters in VMC. This method, based on the Fisher information matrix \cite{ly2017tutorial}, significantly improved the accuracy and efficiency of VMC.

Entering the 21st century, with the rapid development of computational technology, VMC methods combined with modern machine learning techniques opened new research directions. In 2017, Giuseppe Carleo and Matthias Troyer demonstrated the potential of deep learning techniques in optimizing VMC methods by using neural network-based variational wave functions to solve quantum many-body problems \cite{doi:10.1126/science.aag2302}. In the traditional context, machine learning means inputting data to obtain certain patterns, which is the approach of supervised and unsupervised learning. However, in variational Monte Carlo, when solving for the ground state, we do not have any pre-existing ground state wave function data, and can only rely on the variational principle to optimize the variational wave function to minimize the expected value of energy. This method of training neural networks through certain criteria rather than datasets is referred to as \textbf{reinforcement learning}. In the variational Monte Carlo method we constructed next, this method is used to train neural quantum states, which are neural network representations of wave functions.

In this note, variational Monte Carlo method based on neural quantum states is built for spin systems, following the work of Giuseppe Carleo and Matthias Troyer. In traditional variational Monte Carlo, various correlation factors are usually used to introduce strong correlation effects, with a typical example being the Jastrow correlation factor. In fact, a wave function is merely a mapping from the configuration of the system to complex numbers. By setting the output layer of the neural network as a complex neuron, a wave function can be constructed. Using a neural network as the wave function allows for a more generalized expression of various types of interactions, including highly non-local interactions, which are closely related to its non-linear activation functions. Additionally, neural networks can represent relatively complex wave functions with relatively small computational resources when dealing with higher-dimensional systems, which is undoubtedly a "flattening" advantage. In quantum-state tomography \cite{Torlai2018}, the representation method of neural quantum states has already achieved significant results, hinting at its potential in handling larger-sized systems.
\end{abstract}


\end{frontmatter}

\section{Overview of the Variational Monte Carlo Method}
\label{sec1}
\subsection{Origin of Monte Carlo Method}
\label{subsec1.1}
At the beginning of this chapter, we introduce the origin of the idea behind the Variational Monte Carlo (VMC) method. For many-body systems, we denote a specific configuration by $\ket{x}$, and the ground state by $\ket{\phi}$. Then, the wave function of the system can be expressed as a superposition of several configurations, that is,
\begin{equation}
    \ket{\Psi}=\underset{x}{\sum}\langle x|\Psi\rangle\ket{x}
    \label{1}
\end{equation}
Here, we use the completeness of the basis $\{\ket{x}\}$ of the many-body quantum state. Therefore, for a system in the state $\ket{\Psi}$, its observable can be expressed as
\begin{equation}
\langle\hat{O}\rangle=\frac{\langle\Psi|\hat{O}| \Psi\rangle}{\langle\Psi \mid \Psi\rangle}=\frac{\sum_x\langle\Psi \mid x\rangle\langle x|\hat{O}| \Psi\rangle}{\sum_x\langle\Psi \mid x\rangle\langle x \mid \Psi\rangle}=\frac{\sum_x \frac{\langle x|\hat{O}| \Psi\rangle}{\langle x \mid \Psi\rangle}\langle\Psi \mid x\rangle\langle x \mid \Psi\rangle}{\sum_x\langle\Psi \mid x\rangle\langle x \mid \Psi\rangle}=\frac{\sum_x \frac{\langle x|\hat{O}| \Psi\rangle}{\langle x \mid \Psi\rangle}|\langle x \mid \Psi\rangle|^2}{\sum_x|\langle x \mid \Psi\rangle|^2} 
\label{2}
\end{equation}
Thus, we can define a physical quantity,
\begin{equation}
    O_{loc}(x)=\frac{\langle x|\hat{O}| \Psi\rangle}{\langle x |\Psi\rangle}
    \label{3}
\end{equation}
which is called the "local value" of the observable. Therefore, we can reinterpret the expectation value of the observable - $|\langle x|\Psi\rangle|^2$ is the probability of finding the system in the state $\ket{x}$, and the expectation value of the observable can be rewritten as
\begin{equation}
\langle\hat{O}\rangle=\frac{\sum_x O_{\mathrm{loc}}(x)|\langle x \mid \Psi\rangle|^2}{\sum_x|\langle x \mid \Psi\rangle|^2}
\label{4}
\end{equation}
It can be seen that if one wants to directly obtain the expectation value of an observable in this way, it is necessary to obtain all configurations of the system and sum the results under each configuration. However, for many-body systems, it is unrealistic to calculate larger lattice models in this way. For example, without using any symmetry to reduce the state space, the Hilbert space dimension for a spin chain of length $L$ is $2^L$; for fermionic systems, the state space dimension of a one-dimensional fermionic chain of length $L$ is $4^L$. Since the state space dimension of the system grows exponentially with the size of the system, in dealing with many-body problems, people have thought of using reasonable random sampling to solve various problems, which is the Monte Carlo method. We denote the configurations randomly selected as $\mathcal{S}_1=\{x_1,x_2,...,x_N\}$, which is a multisubset of the state space of the system. The expectation value of the observable calculated using them is
\begin{equation}
    \langle\hat{O}\rangle \approx \frac{\sum_{x \in \mathcal{S}_1} O_{\mathrm{loc}}(x)|\langle x \mid \Psi\rangle|^2}{\sum_{x \in \mathcal{S}_1}|\langle x \mid \Psi\rangle|^2}
    \label{5}
\end{equation}
It should be noted that in $\mathcal{S}_1$, we allow the same configuration to appear multiple times. From \eqref{5}, it can be seen that for each configuration $\ket{x}$, its weight determines its role in the measurement of the observable. If completely random sampling is performed, many unimportant $\ket{x}$ will be collected into $\mathcal{S}_1$, while we hope to select $\ket{x}$ that have a large overlap with $\ket{\Psi}$ as much as possible. We can think about this concern through an example. Imagine we are dealing with the two-dimensional Hubbard model
\begin{equation}
    H=-t\underset{\langle i,j\rangle,\sigma}{\sum}\hat{f}_{i,\sigma}^{\dagger}\hat{f}_{j,\sigma}+U\underset{i}{\sum}\hat{n}_{i,\uparrow}\hat{n}_{i,\downarrow}
    \label{6}
\end{equation}
where the fermion number operator $\hat{n}_{i,\sigma}=\hat{f}_{i,\sigma}^{\dagger}\hat{f}_{i,\sigma}$, and $\langle i,j\rangle$ represents a pair of nearest neighbor lattice sites. When the scale of the repulsive interaction $U$ is large, the probability of two fermions occupying the same site (double occupancy) is small. At this time, if completely random sampling is performed, about one-fourth of the configurations will belong to the double occupancy case, which is very unfavorable for the calculation of the expectation value.

To address this issue, the \textbf{importance sampling} method was proposed. When randomly generating configurations, the probability of generating various configurations is equal, but a weight $|\langle x|\Psi\rangle|^2$ needs to be added. In importance sampling, the probability of generating various configurations is determined by the probability $|\langle x|\Psi\rangle|^2$, but the weight of each configuration is equal when calculating the final expectation value. In other words, defining a multisubset of the state space $S_{\rho}$ composed of $N$ configurations generated according to the probability distribution $\rho\propto |\langle x|\Psi\rangle|^2$, the expectation value of the observable is
\begin{equation}
    \langle\hat{O}\rangle \approx \frac{1}{N} \sum_{x \in \mathcal{S}_\rho} O_{\mathrm{loc}}(x)
    \label{7}
\end{equation}
where the denominator $N$ comes from the sum of $N$ ones. $\hat{O}$ in \eqref{4}, \eqref{5}, \eqref{7} can be any observable, such as the Hamiltonian. For a many-body system, people usually focus on its ground state properties. For VMC, we can set a trial wave function $\ket{\Psi(\alpha)}$ that depends on a series of parameters $\{\alpha_i\}$, and minimize the expectation value of the Hamiltonian by changing these parameters, making the trial wave function as close as possible to the ground state of the system. Therefore, the tasks of VMC can be summarized as follows:
\begin{itemize}
    \item Find appropriate parameters $\{\alpha_i\}$ to minimize the expectation value of the Hamiltonian of the trial wave function.
    \item Use the optimized trial wave function obtained in the above step, combined with \eqref{7}, to calculate various physical quantities.
\end{itemize}
Therefore, it can be seen that how to perform importance sampling is an important issue. One common method is to use the Markov chain, which will be discussed in \ref{subsec1.2}.

\subsection{Importance Sampling Using Markov Chains}
\label{subsec1.2}
In this subsection, we mainly discuss how to use \textbf{Markov chains} to generate the state space subset $S_{\rho}$ needed in \eqref{7}. In general, a Markov process is a random operation that can transition the system from one state to another. Its randomness lies in the fact that only the transition probabilities between different states are specified; when the same Markov process is applied to the same state of the system, the results may differ. A Markov chain is a sequence of system states generated by a Markov process. In summary, a Markov process has no "memory," and its transition probabilities to various states only depend on the current state of the system and have no relation to the states the system has previously occupied.

Regarding the reasons for choosing Markov chains, we also provide a brief explanation here. The most notable reason is the convenience of the \textbf{Markov process}, which allows the state space of the system to be traversed within a finite number of steps. Secondly, our goal is to generate a series of configurations that have a large overlap with the wave function $\ket{\Psi}$. If we find such a configuration $\ket{x}$, then a configuration $\ket{x^{\prime}}$ that is not much different from $\ket{x}$ should also have a large overlap with the wave function. If we can find a Markov process that generates a series of configurations $\ket{x^{\prime}}$ through relatively small changes to the existing configuration $\ket{x}$, and these configurations follow a probability density distribution $\rho\propto |\langle x|\Psi\rangle|^2$, then we can use \eqref{7} to calculate the observables. Later, we will introduce one of the most common Markov processes, the \textbf{Metropolis algorithm} combined with local updates.

Next, we will derive the corresponding Metropolis algorithm using the probability distribution $\rho(x)\propto |\langle x|\Psi\rangle|^2$ that we need. Let $T(x\rightarrow x^{\prime})$ be the transition probability from $\ket{x}$ to $\ket{x^{\prime}}$, then by the normalization of probabilities, we have
\begin{equation}
    \underset{x\rightarrow x^{\prime}}{\sum}T(x\rightarrow x^{\prime})=1
    \label{8}
\end{equation}
where $x^{\prime}$ can also be $x$, meaning that the system can remain in its current state with some probability. For a Markov random process, its master equation is
\begin{equation}
    P_{k+1}(x)=P_k(x)+\sum_{x^{\prime}}\left(P_k\left(x^{\prime}\right) T\left(x^{\prime} \rightarrow x\right)-P_k(x) T\left(x \rightarrow x^{\prime}\right)\right)
    \label{9}
\end{equation}
where $P_k(x)$ is the probability that the system is in the configuration $\ket{x}$ after $k$ Markov processes. According to the requirements of the state space subset $S_{\rho}$ we need to generate, we need to set $P_k(x)=\rho(x)\propto |\langle x|\Psi\rangle|^2$ for all $k$. Observing \eqref{9}, we can see that we first need the sum part to be zero, which means
\begin{equation}
    \frac{\rho\left(x^{\prime}\right)}{\rho(x)}=\frac{|\langle x^{\prime}|\Psi\rangle|^2}{|\langle x|\Psi\rangle|^2}=\frac{T\left(x \rightarrow x^{\prime}\right)}{T\left(x^{\prime} \rightarrow x\right)}
    \label{10}
\end{equation}
This is also known as detailed balance. The next problem is how to find the appropriate transition probability $T(x\rightarrow x^{\prime})$. One solution is to decompose the transition probability into two parts
\begin{equation}
    T(x\rightarrow x^{\prime})=S\left(x \rightarrow x^{\prime}\right) A\left(x \rightarrow x^{\prime}\right)
    \label{11}
\end{equation}
Here, the new configuration $\ket{x^{\prime}}$ is first selected with a "suggestion probability" $S(x\rightarrow x^{\prime})$, and if selected, it is accepted with an "acceptance probability" $A(x\rightarrow x^{\prime})$. For the Metropolis algorithm, the suggestion probability is symmetric, that is,
\begin{equation}
    S(x\rightarrow x^{\prime})=S(x^{\prime}\rightarrow x)
    \label{12}
\end{equation}
Regarding the implementation of \eqref{12}, for example in fermionic systems, a simple method is to randomly select an electron and allow it to hop to a random neighboring lattice site. Also, from \eqref{12}, we have
\begin{equation}
    \frac{\rho\left(x^{\prime}\right)}{\rho(x)}=\frac{S\left(x \rightarrow x^{\prime}\right)}{S\left(x^{\prime} \rightarrow x\right)} \frac{A\left(x \rightarrow x^{\prime}\right)}{A\left(x^{\prime} \rightarrow x\right)}=\frac{A\left(x \rightarrow x^{\prime}\right)}{A\left(x^{\prime} \rightarrow x\right)}
    \label{13}
\end{equation}
Since the acceptance probability must be between 0 and 1, we can normalize $A(x\rightarrow x^{\prime})$ and $A(x^{\prime}\rightarrow x)$ by dividing them by a factor, such as $\rho(x)$ when $\rho(x) > \rho(x^{\prime})$, that is
\begin{equation}
    A\left(x \rightarrow x^{\prime}\right)=\left\{\begin{array}{ll}
\frac{\rho\left(x^{\prime}\right)}{\rho(x)} & \text { if } \rho\left(x^{\prime}\right)<\rho(x) \\
1 & \text { otherwise }
\end{array}=\min \left(1, \frac{\rho\left(x^{\prime}\right)}{\rho(x)}\right)\right.
\label{14}
\end{equation}
The suggestion probability defined in \eqref{12} and the acceptance probability defined in \eqref{14} together define the Metropolis algorithm.

In practical calculations, the Markov process needs to run several times to reach the desired probability distribution, which is to reach "equilibrium". Re-examining \eqref{9}, we can see that only when $P_k(x)$ has reached the distribution $P_k(x)=\rho(x)=|\langle x|\Psi\rangle|^2$, will $P_{k+1}(x)$ reach $P_{k+1}(x)=\rho(x)$. However, before the first Markov process is carried out, the initial state of the Markov chain is actually a specific configuration $\ket{y}$, that is, the probability that the system is in the configuration $\ket{x}$ after 0 Markov processes is $P_0(x)=\delta_{x,y}$, which obviously does not meet our requirements. Fortunately, ergodicity and detailed balance are sufficient for the Markov process to converge to a specific distribution that is independent of the initial state \cite{olle2000finite}. Therefore, in actual calculations, the initial results of a Markov chain need to be discarded, that is, only the results after the system has reached equilibrium are valid.

\subsection{Stochastic Reconfiguration}
\label{subsec1.3}
One of the main objectives of VMC is to find the best variational wave function that is closest to the ground state. One common method for finding the ground state wave function is called \textbf{Stochastic Reconfiguration} \cite{PhysRevLett.80.4558}. Let $\ket{\Psi}=\ket{\Psi(\vec{\alpha})}$ be the trial wave function, which depends on a series of parameters $\vec{\alpha}=(\alpha_1,\alpha_2,...,\alpha_p)$. In the process of optimizing the trial wave function, we adjust these parameter values to achieve the lowest possible energy. For the updated wave function $\ket{\Psi^{\prime}}$, we can make the following ansatz,
\begin{equation}
\left|\Psi^{\prime}\right\rangle=\delta \alpha_0|\Psi\rangle+\sum_{k^{\prime}=1}^p \delta \alpha_{k^{\prime}} \frac{\partial}{\partial \alpha_{k^{\prime}}}|\Psi\rangle
\label{15}
\end{equation}
which is essentially the Taylor expansion of the wave function $\ket{\Psi}$ with respect to the components of $\vec{\alpha}$. Next, we need to find a series of $\delta \alpha_i$ that can lower the energy. By inserting the completeness relation of the basis composed of all configurations of the system into \eqref{15}, we get
\begin{equation}
\begin{aligned}
\left|\Psi^{\prime}\right\rangle & =\delta \alpha_0|\Psi\rangle+\sum_{k^{\prime}=1}^p \delta \alpha_{k^{\prime}} \frac{\partial}{\partial \alpha_{k^{\prime}}} \sum_x|x\rangle\langle x \mid \Psi\rangle \\
& =\delta \alpha_0|\Psi\rangle+\sum_{k^{\prime}=1}^p \delta \alpha_{k^{\prime}} \sum_x \frac{\partial\langle x \mid \Psi\rangle}{\partial \alpha_{k^{\prime}}}|x\rangle \\
& =\delta \alpha_0|\Psi\rangle+\sum_{k^{\prime}=1}^p \delta \alpha_{k^{\prime}} \sum_x \frac{\partial \log \langle x \mid \Psi\rangle}{\partial \alpha_{k^{\prime}}}\langle x \mid \Psi\rangle|x\rangle
\end{aligned}
\label{16}
\end{equation}
Introducing the operator $\hat{\Delta}_k=\begin{cases}\mathbf{1} & \text { for } k=0 \\ \sum_x \frac{\partial \log \langle x \mid \Psi\rangle}{\partial \alpha_k}|x\rangle\langle x| & \text { for } k \neq 0\end{cases}$, then \eqref{16} can be rewritten as
\begin{equation}
\left|\Psi^{\prime}\right\rangle=\delta \alpha_0\left|\Psi\right\rangle+\delta \alpha_k \hat{\Delta}_k\left|\Psi\right\rangle
\label{17}
\end{equation}
With such an ansatz, we can seek the ground state of the system by relating it to \textbf{imaginary time evolution}. For any system, we can write its wave function as a coherent superposition of several eigenstates of the Hamiltonian
\begin{equation}
    \ket{\Psi}=\underset{i}{\sum}c_i(0)\ket{\psi}
    \label{18}
\end{equation}
After applying the time evolution operator, it becomes
\begin{equation}
     \ket{\Psi(t)}=\underset{i}{\sum}c_i(0)e^{-i\hat{H}t}\ket{\psi}=\underset{i}{\sum}c_i(0)e^{-iE_it}\ket{\psi}
     \label{19}
\end{equation}
Now we perform a Wick rotation, which projects the system's evolution time $t$ onto imaginary time $\tau=-it$, yielding
\begin{equation}
     \ket{\Psi(\tau)}=\underset{i}{\sum}c_i(0)e^{-E_i\tau}\ket{\psi}
     \label{20}
\end{equation}
From \eqref{20}, it can be seen that the higher the energy of an eigenstate, the faster it decays with imaginary time evolution. At a certain imaginary time $\tau$, the ground state $\ket{\phi}$ will dominate the linear combination, which is why we can use imaginary time evolution to find the ground state of the system. Therefore, each ansatz step $\ket{\Psi^{\prime}}$ should correspond to one step of imaginary time evolution. That is, for imaginary time evolution
\begin{equation}
|\Psi^{''}\rangle=U(\tau)\left|\Psi\right\rangle=e^{-\tau \hat{H}}\left|\Psi\right\rangle \sim(\mathbb{I}-\tau \hat{H})\left|\Psi\right\rangle
\label{21}
\end{equation}
it needs to correspond to \eqref{17}, and the criterion for determining the correspondence is to minimize the norm of the difference between the ansatz and the imaginary time evolution, i.e., minimizing the norm of $\ket{\Psi^{\prime}}-\ket{\Psi^{''}}$. Thus, the above problem can be transformed into
\begin{equation}
\ket{\Psi^{\prime}}=\ket{\Psi^{''}}\Rightarrow (\delta \alpha_0+\overunderset{p}{k=1}{\sum}\delta \alpha_k\hat{\Delta}_k)\ket{\Psi}=(1-\hat{H}\tau)\ket{\Psi}
\label{22}
\end{equation}
Projecting \eqref{22} onto $\hat{\Delta}^{\dagger}_{k^{\prime}}\ket{\Psi}$, we get a system of linear equations for $\delta \vec{\alpha}$,
\begin{equation}
    \sum_{k=0}^p \delta \alpha_{k^{\prime}}\left\langle\Psi\left|\hat{\Delta}_{k^{\prime}} \hat{\Delta}_{ k}\right| \Psi\right\rangle=\left\langle\Psi\left|\hat{\Delta}_{k^{\prime}}(1-\tau\hat{H})\right| \Psi\right\rangle \quad  k^{\prime} \in\{0, \ldots, p\}
    \label{23}
\end{equation}
Using the shorthand notation $\langle...\rangle$ for the expectation value with respect to $\ket{\Psi}$, \eqref{23} can be simplified to
\begin{equation}
    \sum_{k=0}^p \delta \alpha_{k^{\prime}}\langle\hat{\Delta}_{k^{\prime}} \hat{\Delta}_{ k}\rangle=\langle\hat{\Delta}_{k^{\prime}}(1-\tau\hat{H})\rangle \quad k^{\prime} \in\{0, \ldots, p\}
    \label{24}
\end{equation}
Setting $k^{\prime}=0$, since $\hat{\Delta}_{k^{\prime}}=1$, from \eqref{24} we get
\begin{equation}
    \delta \alpha_0=1-\tau\langle \hat{H}\rangle-\sum_{k=1}^p\delta \alpha_k \langle \hat{\Delta}_k\rangle
    \label{25}
\end{equation}
Substituting \eqref{25} into \eqref{24}, we have
\begin{equation}
    \sum_{k=1}^p\delta \alpha_k(\langle \hat{\Delta}_{k^{\prime}}\hat{\Delta}_k\rangle-\langle \hat{\Delta}_k\rangle\langle \hat{\Delta}_{k^{\prime}}\rangle)=\tau (\langle \hat{\Delta}_{k^{\prime}}\rangle\langle \hat{H}\rangle-\langle \hat{\Delta}_{k^{\prime}}\hat{H}\rangle) \quad k^{\prime} \in\{0, \ldots, p\}
    \label{26}
\end{equation}
In \eqref{26}, all the expectation values can be calculated using \eqref{7}. Here we can simply calculate the local values of several physical quantities,
\begin{equation}
    \begin{aligned}
        \Delta_{k, \text { loc }}(x)&=\frac{\left\langle x\left|\hat{\Delta}_{ k}\right| \Psi\right\rangle}{\langle x \mid \Psi\rangle}=\frac{\left\langle x\left|\sum_{x^{\prime}} \frac{\partial \log \left\langle x^{\prime} \mid \Psi\right\rangle}{\partial \alpha_k}\right| x^{\prime}\right\rangle\left\langle x^{\prime} \mid \Psi\right\rangle}{\langle x \mid \Psi\rangle}=\frac{\partial \log \langle x \mid \Psi\rangle}{\partial \alpha_k}\\
\frac{\left\langle x\left|\hat{\Delta}_{k} \hat{\Delta}_{k^{\prime}}\right| \Psi\right\rangle}{\langle x \mid \Psi\rangle} & =\frac{\left\langle x\left|\sum_{x^{\prime}} \frac{\partial \log \left\langle x^{\prime} \mid \Psi\right\rangle}{\partial \alpha_k}\right| x^{\prime}\right\rangle\left\langle x^{\prime}\left|\sum_{x^{\prime \prime}} \frac{\partial \log \left\langle x^{\prime \prime} \mid \Psi\right\rangle}{\partial \alpha_{k^{\prime}}}\right| x^{\prime \prime}\right\rangle\left\langle x^{\prime \prime} \mid \Psi\right\rangle}{\langle x \mid \Psi\rangle} \\
& =\frac{\partial \log \langle x \mid \Psi\rangle}{\partial \alpha_k} \frac{\partial \log \langle x \mid \Psi\rangle}{\partial \alpha_{k^{\prime}}}=\Delta_{ k, \text { loc }}(x) \Delta_{ k^{\prime}, \text { loc }}(x) \\
\frac{\left\langle x\left|\hat{\Delta}_{ k} \hat{H}\right| \Psi\right\rangle}{\langle x \mid \Psi\rangle} & =\sum_{x^{\prime}}\left\langle x\left|\hat{\Delta}_{ k} \hat{H}\right| x^{\prime}\right\rangle \frac{\left\langle x^{\prime} \mid \Psi\right\rangle}{\langle x \mid \Psi\rangle} \\
& =\sum_{x^{\prime} x^{\prime \prime}}\left\langle x\left|\frac{\partial \log \left\langle x^{\prime \prime} \mid \Psi\right\rangle}{\partial \alpha_{k^{\prime}}}\right| x^{\prime \prime}\right\rangle\left\langle x^{\prime \prime}|\hat{H}| x^{\prime}\right\rangle \frac{\left\langle x^{\prime} \mid \Psi\right\rangle}{\langle x \mid \Psi\rangle} \\
& =\frac{\partial \log \langle x \mid \Psi\rangle}{\partial \alpha_k} \frac{\langle x|\hat{H}| \Psi\rangle}{\langle x \mid \Psi\rangle}=\Delta_{ k, \text { loc }}(x) E_{\mathrm{loc}}(x)
    \end{aligned}
    \label{27}
\end{equation}
By this, the linear relationship of $\delta \vec{\alpha}$ in \eqref{26} is determined. Writing it in a more compact form, we get
\begin{equation}
    \begin{aligned}
\boldsymbol{S} \delta \vec{\alpha}=\tau\vec{f} \quad  \quad S_{k k^{\prime}} & =\left\langle\hat{\Delta}_{ k} \hat{\Delta}_{ k^{\prime}}\right\rangle-\left\langle\hat{\Delta}_{ k}\right\rangle\left\langle\hat{\Delta}_{ k^{\prime}}\right\rangle \\
f_k & =\left\langle\hat{\Delta}_{k}\right\rangle\langle\hat{H}\rangle-\left\langle\hat{\Delta}_{ k} \hat{H}\right\rangle
\end{aligned}
\label{28}
\end{equation}
It is worth noting that $\boldsymbol{S}$ in \eqref{28} is generally referred to as the \textbf{quantum geometric tensor}. Since the quantum geometric tensor is calculated using the Monte Carlo method, it often has small eigenvalues tending to 0, leading to numerical instability. Therefore, we usually need to add a diagonal shift to the quantum geometric tensor, adding a small constant $\varepsilon \sim 10^{-5}-10^{-2}$ to all its diagonal elements.

After solving for $\delta \vec{\alpha}$ in \eqref{28}, we can update the parameters $\vec{\alpha}=(\alpha_1,...,\alpha_p)$ accordingly,
\begin{equation}
\vec{\alpha}^{\prime}=\vec{\alpha}+\lambda \delta \vec{\alpha}
\label{29}
\end{equation}
where the step size $\lambda$ should be chosen appropriately: it needs to be large enough to ensure rapid convergence to the ground state, but small enough to ensure the stability of the algorithm. Generally, a good approach is to start with a larger $\lambda$ and gradually decrease it.

\section{Nural Quantum States}
\label{sec2}
\subsection{Introduction to NQS}
\label{subsec2.1}
The neural quantum state (NQS) \cite{vivas2022neuralnetwork} is essentially a representation of the wave function, expected to be used in the ansatz part of the VMC method discussed earlier. Consider a many-body quantum state with $N$ discrete degrees of freedom $S=(S_1,S_2,...,S_N)$. Whether it is spin, fermion occupancy number, or even boson occupancy number, the many-body wave function is simply a mapping from $S$ to a complex number. Therefore, as long as the output layer of the neural network is set to the real and imaginary parts, or the amplitude and phase parts of the wave function, it can theoretically be trained to represent the many-body wave function.

In the introduction of the Monte Carlo method in \ref{subsec1.1}, we considered that for a many-body system, the dimension of its state space typically grows exponentially with the system size. However, the information needed to describe a particular quantum state is much less than the dimension of the system's state space. Algorithms that rely on the wave function itself and can solve many-body problems using relatively small amounts of information can be broadly divided into two categories: sampling-based methods and compression-based methods. The most typical sampling-based method is the quantum Monte Carlo (QMC) method, which aims to calculate various expectation values by extracting a series of valuable configurations. Compression-based methods, which aim to represent the wave function with less memory, mainly include matrix product states (MPS) \cite{_stlund_1995} and the more general tensor networks \cite{biamonte2020lectures}. However, both methods have situations that are difficult or impossible to handle: for QMC, it is the notorious sign problem, and for compression-based methods, it is the quantum state of high-dimensional systems. In scenarios where common numerical methods are ineffective, neural networks provide an alternative option.

\subsection{Working Principle and Basic Architectures of Neural Networks}
\label{subsec2.2}
Before introducing neural quantum states, we first need to understand the working principles and some basic architectures of neural networks \cite{10.21468/SciPostPhysLectNotes.29}. The purpose of a neural network, simply put, is to use a series of nonlinear functions to generate a function $y=F_{\theta}(x)$ (where $\theta$ are the adjustable parameters) to fit any smooth function $y=F(x)$. For a real dataset, $y=F(x)$ is often unknown and is obtained by fitting any function through a series of inputs $x$ and outputs $y$ in the dataset using the neural network.

\subsubsection{Neurons and Their Interactions}
\label{subsubsec2.2.1}
The basic unit of a neural network is a neuron, which has a real scalar value $y$. This scalar value comes from the scalar values $y_k$ of all the neurons that feed into this neuron. The process can be summarized as follows:
\begin{itemize}
    \item Calculate a linear function $z=\sum_k w_ky_k+b$, where the matrix $w$ is called the weights and the vector $b$ is called the bias. These two parameters are members of the aforementioned parameters $\theta$.
    \item Use a nonlinear function $f(z)$, usually the sigmoid function or the ReLU function, to obtain the value of the neuron $y=f(z)$. The sigmoid function is a smoothed step function $f(x)=1 /\left(1+e^{-x}\right)$, and the ReLU function is a piecewise function $f(x)=0\quad x<0$ \& $f(x)=x\quad x \geq 0$.
\end{itemize}
\begin{figure}
    \centering
    \subfloat[Sigmoid]{
        \includegraphics[width=0.5\textwidth]{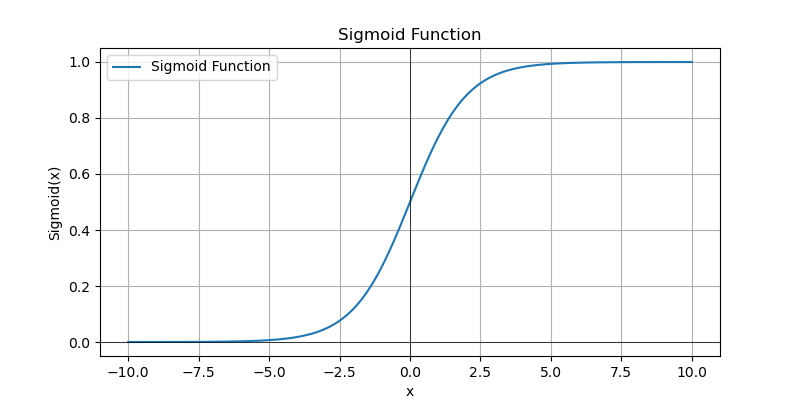}
    }
    \subfloat[ReLU]{
        \includegraphics[width=0.5\textwidth]{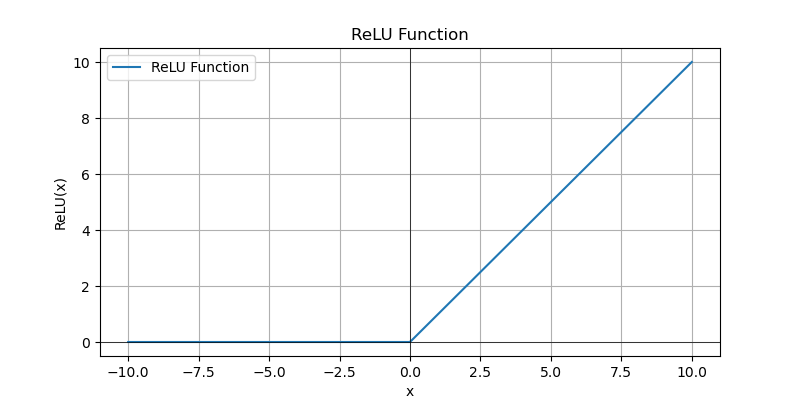}
    }
    \caption{Two common activation functions are: (a) the Sigmoid function and (b) the ReLU (rectified linear unit) function. Both of them take values between 0 and 1.}
    \label{activation_functions}
\end{figure}
To clarify the notation in multi-layer neural networks, we use $y_k^{(n)}$ to denote the value of the $k$-th neuron in the $n$-th layer. Therefore, the component $w_{j k}^{(n+1)}$ of the weights tells us how the $k$-th neuron in the $n$-th layer affects the $j$-th neuron in the $(n+1)$-th layer. Thus, the value of the $j$-th neuron in the $(n+1)$-th layer can be determined by the following equation:
\begin{equation}
    \begin{gathered}
z_j^{(n+1)}=\sum_k w_{j k}^{(n+1)} y_k^{(n)}+b_j^{(n+1)} \\
y_j^{(n+1)}=f\left(z_j^{(n+1)}\right)
\end{gathered}
\label{30}
\end{equation}
For the entire neural network, its output is calculated layer by layer starting from the 0th layer ($n=0$), which is the layer of input data $x$, until the final output (which can be a single $y$ value or a vector composed of multiple $y$ values) is obtained. This output is compared with the $y$ value provided in the dataset to continuously optimize the weights and biases in the neural network. However, it is obvious that larger neural networks will occupy more memory. For example, if the $n$-th layer has $N^{(n)}$ neurons and the $(n+1)$-th layer has $N^{(n+1)}$ neurons, the dimension of the weight matrix representing the influence of the $n$-th layer on the $(n+1)$-th layer is $N^{(n+1)}\times N^{(n)}$.

\subsubsection{Stochastic Gradient Descent Method}
\label{subsubsec2.2.2}
To evaluate the performance of a neural network and optimize its parameters, we need to define a measurable quantity to quantify the deviation between the neural network's output and the target function. This quantity is the cost function. The simplest cost function is the quadratic deviation between the neural network's output $F_{\theta}(x)$ and the target function $F(x)$. First, we can define a loss function for a specific sample (specific dataset input $x$)
\begin{equation}
    C_x(\theta)=\left|F_\theta(x)-F(x)\right|^2
    \label{31}
\end{equation}
Averaging the loss function over all samples gives the cost function
\begin{equation}
    C(\theta)=\left\langle C_x(\theta)\right\rangle_x
    \label{32}
\end{equation}
The core task of machine learning is to minimize the cost function. Therefore, defining the cost function is as important as defining the Hamiltonian or Lagrangian in solving physical problems. One basic idea is to use stochastic gradient descent (SGD) \cite{Rosenblatt1958ThePA} in the linear space spanned by the parameters $\theta$ to find the minimum.
The idea of gradient descent is straightforward: simply descend along the gradient direction of the cost function with respect to the parameters $\theta$, that is,
\begin{equation}
    \delta \theta_k=-\eta \frac{\partial C(\theta)}{\partial \theta_k}
    \label{33}
\end{equation}
where the parameter $\eta$ is called the \textbf{learning rate}. This parameter directly determines the final effectiveness of the neural network. If the learning rate is too small, the optimization process will be too slow; if the learning rate is too large, the optimal solution may be overlooked. However, the gradient descent method based on \eqref{33} has two significant issues:
\begin{itemize}
    \item It requires averaging over all possible dataset inputs $x$, which is computationally expensive.
    \item The process of calculating the gradient itself involves many elements of the parameters $\theta$, so we need a more efficient way to calculate the gradient.
\end{itemize}
To address the first issue, we use an approach similar to the basic idea of Monte Carlo: randomly sample a portion of the training samples (called a batch) to average,
\begin{equation}
    C(\theta) \approx \frac{1}{N} \sum_{j=1}^N C_{x_j}(\theta) \equiv\left\langle C_x(\theta)\right\rangle_{\text {batch }}
    \label{34}
\end{equation}
The batch average defines the improved stochastic gradient descent method based on the gradient descent method,
\begin{equation}
    \delta \theta_k=-\eta\left\langle\frac{\partial C_x(\theta)}{\partial \theta_k}\right\rangle_{\text {batch }}=-\eta \frac{\partial C(\theta)}{\partial \theta_k}+\Omega
    \label{35}
\end{equation}
where $\Omega$ represents the noise term, which is usually expected to average to zero after several steps of gradient descent. In summary, the stochastic gradient descent method is suitable for cases with a smaller learning rate.

To address the second issue of low efficiency in calculating the derivatives of a large number of parameters, we can use the chain rule, also known as backpropagation. For a quadratic loss function, we have
\begin{equation}
    \frac{\partial C_x(\theta)}{\partial \theta_k}=2 \sum_l\left(\left[F_\theta(x)\right]_l-[F(x)]_l\right) \frac{\partial\left[F_\theta(x)\right]_l}{\partial \theta_k}
    \label{36}
\end{equation}
where the sum over $l$ represents the neurons, and $\left[F_\theta(x)\right]_l=y_l^{(n)}$ represents the value of the $l$-th neuron in the $n$-th layer. According to \eqref{30}, we can write the gradient with respect to the parameters $\theta$ as
\begin{equation}
    \frac{\partial y_l^{(n)}}{\partial \theta_k}=f^{\prime}\left(z_l^{(n)}\right) \frac{\partial z_l^{(n)}}{\partial \theta_k}=f^{\prime}\left(z_l^{(n)}\right)\sum_m w_{l m}^{(n, n-1)} \frac{\partial y_m^{(n-1)}}{\partial \theta_k}
    \label{37}
\end{equation}
In the case of \eqref{37}, the parameter $\theta_k$ does not include the weights and biases between the $n-1$ and $n$ layers. This is a recursive relation and can also be written in the form of matrix and vector multiplication. Thus, we define the following matrix
\begin{equation}
    M_{l m}^{(n, n-1)}=w_{l m}^{(n, n-1)} f^{\prime}\left(z_m^{(n-1)}\right)
    \label{38}
\end{equation}
Then, according to the calculation of $\frac{\partial z_l^{(n)}}{\partial \theta_k}$ in \eqref{37}, we have
\begin{equation}
    \frac{\partial z_l^{(n)}}{\partial \theta_k}=\left[M^{(n, n-1)} M^{(n-1, n-2)} \ldots M^{\left(n^{\prime}+1, n^{\prime}\right)} \frac{\partial z^{\left(n^{\prime}\right)}}{\partial \theta_k}\right]_l
    \label{39}
\end{equation}
Similarly, $\theta_k$ cannot be the weights and biases between the $n$-th and $n^{\prime}$-th layers. Based on the above considerations, we can introduce the backpropagation method, which can be summarized as follows:
\begin{itemize}
    \item In the output layer (set as the $N$-th layer here), define a "deviation vector" $\Delta_j=(y_j^N-[F(x)]_j)f^{\prime}\left(z_j^{(N)}\right)$.
    \item For each layer starting from $n=N$, store the derivatives related to the weights and biases in that layer $\frac{\partial C_x(\theta)}{ \partial \theta_k}=\Delta_j \frac{\partial z_i^{(n)}} {\partial \theta_k}$, where $\theta_k$ appears as an explicit variable in $z_i^{(n)}$.
    \item Move one layer toward the input layer and set the new deviation vector $\Delta_j^{(\text {new })}=\sum_k \Delta_k M_{k j}^{(n, n-1)}$. Repeat all the above steps until all the derivatives are solved.
\end{itemize}
This algorithm requires computational resources comparable to the process of calculating the output of a neural network for a given input, which is called the forward pass. It can also utilize the values of each neuron calculated during the forward pass. This efficient gradient calculation method makes the operation of neural networks a feasible procedure, laying the foundation for subsequent applications and developments.

These are the basic working principles of neural networks. In addition, how to construct neural networks is also an important topic. In the following sections, we will discuss the construction of several typical neural networks, such as convolutional neural networks (CNNs) \cite{cireşan2012multicolumn,Fukushima1980,LeCun2015}, Boltzmann machines \cite{Ackley1985ALA,PhysRevLett.35.1792,doi:10.1073/pnas.79.8.2554}, and their applications in computing specific strongly correlated systems.

\subsubsection{Convolutional Neural Networks}
\label{subsubsec2.2.3}
In the previous subsection, when discussing the working principles of neural networks, we assumed that the neural network is a dense neural network, meaning that every neuron in adjacent layers is connected. In contrast, convolutional neural networks (CNNs) are sparse neural networks, as we will see that their weight matrices are sparse matrices. CNNs exhibit translational symmetry by constraining the weights of the neural network. The term "convolution" is used because, when dealing with linear responses, we have
\begin{equation}
    A(x)=\int G\left(x-x^{\prime}\right) F\left(x^{\prime}\right) d x^{\prime}
    \label{40}
\end{equation}
We can see that the response at point $x$ corresponding to point $x^{\prime}$ depends only on the distance between them. If both are shifted by the same amount, the final calculated response remains unchanged, which is translationally symmetric, and the mathematical form of \eqref{40} is precisely convolution.

The weights in a convolutional neural network also depend only on the "distance" between two neurons. Considering the one-dimensional case, we have
\begin{equation}
    w_{i j}^{(n+1, n)}=w^{(n+1, n)}(i-j)
    \label{41}
\end{equation}
Equation \eqref{41} is also known as the "convolution kernel." For the kernel, we need to truncate it, setting the corresponding weight elements to zero for $|i-j| > d$. This significantly reduces the number of non-zero elements in the weights. For example, if both adjacent layers have $M$ neurons, the dimension of the weight matrix is $M \times M$, with $M^2$ non-zero elements in the unconstrained case. After truncation, a neuron in the $(n+1)$-th layer has contributions from only $2d+1$ neurons in the $n$-th layer, so its corresponding weight is a vector with only $2d+1$ components.
\begin{figure}
    \centering{
        \includegraphics[width=0.75\textwidth]{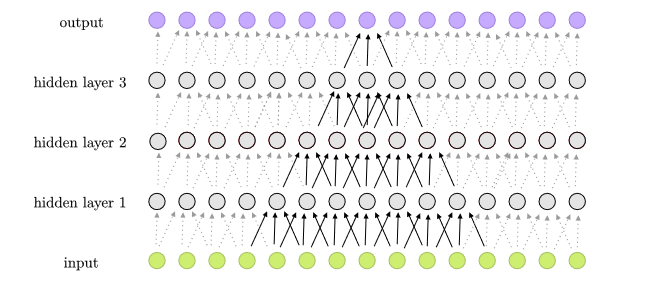}
    }
    \caption{The working architecture of convolutional neural networks \cite{oshea2015introduction}: The receptive field of the convolutional neural network includes the nearest and next-nearest neurons in the previous layer, without applying periodic boundary conditions. It can be seen that the connections between neurons in convolutional neural networks are much sparser compared to dense neural networks, allowing for larger numbers of neurons per layer and more layers, which is crucial in extracting image features.}
    \label{CNN}
\end{figure}
It should be noted that the weights corresponding to each neuron in a convolutional neural network are "shared." This means that for each neuron at position $i$ in the $(n+1)$-th layer, we can define a local neighborhood in the $n$-th layer that satisfies $|j-i| \leq d$, and use the values of the neurons in this local neighborhood (receptive field) as inputs, with the same convolution kernel as weights, to obtain $y_i^{(n+1)}$, the value of the $i$-th neuron in the $(n+1)$-th layer. Often, the receptive field of a neuron extends beyond the boundaries of the previous layer. In this case, different padding strategies are considered. For periodic boundary conditions, we can use periodic padding, where the values of neurons beyond the boundary are taken from the other end of the previous layer.

The two-dimensional case for convolutional neural networks is similar. Simply put, the weights form an array with dimensions $(2d+1) \times (2d+1)$, corresponding to each neuron in the receptive field.

Convolutional neural networks not only have translational invariance but also split the transformation in \eqref{30} into multiple channels. Using $c$ and $c^{\prime}$ as channel indices, the indices of neurons are determined by their position $i$ in a certain layer and the channel index, $(i,c)$. Thus, the transformation in convolutional neural networks can be rewritten as
\begin{equation}
    z_{(i, c)}^{(n+1)}=\sum_j w_{c c^{\prime}}^{(n+1, n)}(i-j) y_{\left(j, c^{\prime}\right)}^{(n)}+b_c^{(n+1)}
    \label{42}
\end{equation}
In the following chapters, we will discuss how to use convolutional neural networks to compute the two-dimensional $J_1-J_2$ model with strong frustration.

\subsubsection{Restricted Boltzmann Machines}
\label{subsubsec2.2.4}
The basic purpose of a Boltzmann machine is to model the probability distribution $P_0(v)$ of the observed data $v$. In other words, we do not know the form of $P_0(v)$ in advance, but we let the Boltzmann machine approximate it by observing as many samples as possible. The starting point is to set up a statistical model whose Boltzmann distribution can approximate the probability distribution $P_0(v)$. For a Boltzmann distribution, energy plays a key role. Thus, the energy $E_{\theta}$, which depends on a series of parameters $\theta$, is the quantity optimized during the training of the Boltzmann machine.
\begin{figure}
    \centering{
        \includegraphics[width=0.75\textwidth]{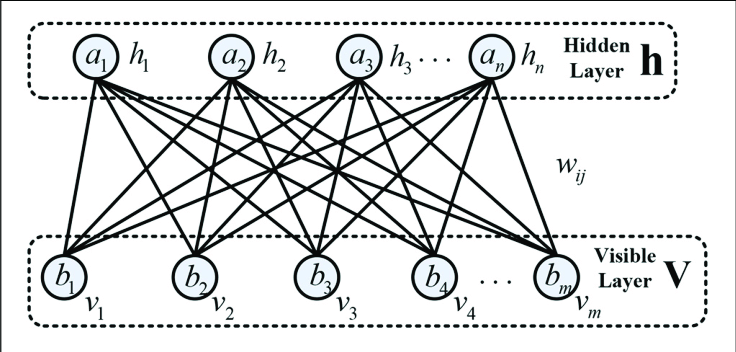}
    }
    \caption{The working architecture of a Boltzmann machine \cite{articleChu}: $v_i$ represents visible layer neurons, $h_i$ represents hidden layer neurons. Each neuron in the visible layer is connected to each neuron in the hidden layer, making the Boltzmann machine a dense network.}
    \label{BM}
\end{figure}
A Boltzmann machine has a layer of visible units $v$ and a layer of hidden units $h$. The Boltzmann distribution we mentioned above is the joint distribution of $(h,v)$:
\begin{equation}
    P(v, h)=\frac{e^{-E(v, h)}}{Z} \quad Z=\sum_{v, h} e^{-E(v, h)}
    \label{43}
\end{equation}
where $Z$ is the partition function. Thus, we can summarize the goal of training a Boltzmann machine as adjusting the parameters $\theta$ to achieve
\begin{equation}
    P_v(v)=\sum_h P(v, h) \approx P_0(v)
    \label{44}
\end{equation}
where $P_v(v)$ is the marginal probability distribution of $v$. A common architecture is the restricted Boltzmann machine (RBM) \cite{Fischer2012}, whose energy is
\begin{equation}
    E(v, h)=-\sum_i a_i v_i-\sum_j b_j h_j-\sum_{i, j} v_i w_{i j} h_j
    \label{45}
\end{equation}
Note that in \eqref{45}, there are no $v-v$ or $h-h$ couplings, so this model is called "restricted." A classic setup is to set the values of $v_i$ and $h_j$ to binary, i.e., 0 or 1. In this case, the model is similar to the Ising model, with the analogous magnetic field terms $a_i$ and $b_j$ also called biases, but the coupling is between any $v_i$ and $h_j$. The parameters $\theta$ of the model are $a_i$, $b_j$, and $w_{ij}$, which are continuously optimized during training. The goal of training is to find a set of parameters $\theta$ that minimize the cost function. For RBMs, we train them to make the marginal probability distribution of the visible layer $P_v(v)$ as close as possible to the target probability distribution $P_0(v)$. Thus, the cost function is
\begin{equation}
    C=-\sum_v P_0(v) \log P_v(v)
    \label{46}
\end{equation}
This is also known as cross-entropy \cite{MacKay2003}, which can only reach its minimum when each $P_v(v)$ exactly equals $P_0(v)$. The detailed derivation can be found in \ref{Minimum Value of Cross Entropy}. Equation \eqref{46} appears simple in form, but in practice, we find that for $N$ visible units $v$, even if their values are only 0 or 1, summing over them requires summing over $2^N$ different results. Fortunately, when performing gradient descent, we mainly consider gradient descent concerning the weight components $w$, and the gradient of the cross-entropy concerning the weights can be solved using sampling. The negative gradient of the cost function concerning the weights is
\begin{equation}
    -\frac{\partial}{\partial w_{i j}} C=\sum_v P_0(v) \frac{\partial}{\partial w_{i j}} \log P_v(v)=\left\langle v_i h_j\right\rangle_{P_0}-\left\langle v_i h_j\right\rangle_{P_v}
    \label{47}
\end{equation}
\begin{equation}
    \begin{aligned}
\left\langle v_i h_j\right\rangle_{P_0} & \equiv \sum_{v, h} v_i h_j P(h \mid v) P_0(v) \\
\left\langle v_i h_j\right\rangle_{P_v} & \equiv \sum_{v, h} v_i h_j P(h \mid v) P_v(v)
\end{aligned}
\label{48}
\end{equation}
where
\begin{equation}
    P(h \mid v)=\frac{P(v, h)}{P_v(v)}
    \label{49}
\end{equation}
is the conditional probability, i.e., the probability distribution of $h$ given $v$. To compute \eqref{47}, we need a method to sample $v$ according to the marginal probability distribution $P_v(v)$ and to sample $h$ given a specific $v$ according to the conditional probability. Generally, sampling according to a specific probability distribution can be done using the Monte Carlo method, constructing a Markov process that satisfies the detailed balance condition. If we want to obtain a Boltzmann distribution, we need to replace the probability distribution $\rho(x)$ in \eqref{10} with the Boltzmann factor $e^{-E_x}$ (where $k_BT=1$), and the remaining details are as discussed in \ref{subsec1.2}.

Of course, the Markov process required in a Boltzmann machine differs somewhat from the one discussed in \ref{subsec1.2}. In a Boltzmann machine, there are visible layers $v$ and hidden layers $h$, which differ. Consider a Markov chain: from a configuration $v$ in the visible layer to a configuration $h$ in the hidden layer, then to a configuration $v^{\prime}$ in the visible layer, and so on. Each step from the visible layer to the hidden layer, or vice versa, corresponds to a transition probability which is the conditional probability $T(v \rightarrow h)=P(h|v)$ or $T(h \rightarrow v^{\prime})=P(v^{\prime}|h)$. Thus, the detailed balance condition for this Markov chain can be written as
\begin{equation}
    \frac{T(v\rightarrow h)}{T(h\rightarrow v)}=\frac{P(h|v)}{P(v|h)}=\frac{P_h(h)}{P_v(v)}
    \label{50}
\end{equation}
After a certain number of Markov processes, this Markov chain will converge to a state where the visible and hidden layers conform to their respective marginal probability distributions.

To specifically implement a Monte Carlo update step, we need to first calculate the conditional probability
\begin{equation}
    P(h \mid v)=\frac{e^{-E(v, h)}}{Z P_v(v)}=\underset{j}{\Pi} \frac{e^{z_j h_j}}{1+e^{z_j}}
    \label{51}
\end{equation}
where
\begin{equation}
    z_j=b_j+\sum_i v_i w_{i j}
    \label{52}
\end{equation}
For details, see \ref{Conditional Probability in Restricted Boltzmann Machines}. It can be seen that for a hidden layer unit, the probability of $h_j=1$ is a sigmoid function. Thus, for RBMs, the Monte Carlo sampling process of the hidden layer state can be roughly divided into the following two steps:
\begin{itemize}
    \item For each hidden layer unit, calculate the probability $\frac{e^{z_j}}{1+e^{z_j}}$, just as we calculate the value of a neuron in a dense neural network.
    \item For each hidden layer unit, assign it a value of 1 with probability $\frac{e^{z_j}}{1+e^{z_j}}$, and a value of 0 with probability $\frac{1}{1+e^{z_j}}$.
\end{itemize}
Then, to perform the Markov process from the hidden layer to the visible layer, replace $z_j$ with
\begin{equation}
    z_i^{\prime}=a_i+\sum_j w_{i j} h_j
    \label{53}
\end{equation}
and replace the conditional probability with
\begin{equation}
    P(v^{\prime} \mid h)=\frac{e^{-E(v^{\prime}, h)}}{Z P_h(h)}=\underset{i}{\Pi} \frac{e^{z^{\prime}_i v^{\prime}_i}}{1+e^{z^{\prime}_i}}
    \label{54}
\end{equation}
Starting from the initial visible layer state $v$, sample the hidden layer state $h$, then sample the visible layer state $v^{\prime}$, completing one Monte Carlo sampling of the RBM.

Finally, we need to measure the quantities in \eqref{48} through Monte Carlo methods to update the weights. It is relatively simple to calculate under the probability distribution $P_0(v)$ because the samples we have in advance are sampled from this distribution. Specifically, $P_0(v)$ is the empirical distribution function of the dataset used at the beginning as the visible layer state $v$. In each Monte Carlo sampling, we obtain a pair $(v,h)$ and its conditional probability $P(h|v)$. If we perform $N$ Monte Carlo samplings, the first equation in \eqref{48} can be written as
\begin{equation}
\left\langle v_i h_j\right\rangle_{P_0} \approx \frac{1}{N} \sum_{v, h} v_i h_j P(h \mid v) P_0(v)
\label{55}
\end{equation}
Here, the sum over $v$ and $h$ no longer means summing over all visible and hidden layer states, but summing over each pair $(v,h)$ obtained from Monte Carlo sampling. Thus, we again use the idea of random sampling to greatly reduce the computational difficulty.

For $P_v(v)$, the system needs to undergo a considerable number of Markov processes before converging to this probability distribution. Hence, the second equation in \eqref{48} seems difficult to calculate. However, if the marginal probability distribution $P_v(v)$ of the RBM is already quite close to $P_0(v)$, then sampling based on $P_0(v)$ will be quite similar to sampling based on $P_v(v)$. Subsequently, starting from this sample $v$, a few more Monte Carlo samplings can be performed to get closer to the probability distribution $P_v(v)$. In practice, a simple method is to add a few more steps: $v \rightarrow h \rightarrow v^{\prime} \rightarrow h^{\prime}$. Here, $v^{\prime}$ can be considered a good approximation of a sample drawn from $P_v(v)$. If we always use this approximation in Monte Carlo sampling, \eqref{47} can be directly rewritten as
\begin{equation}
    \left\langle v_i h_j\right\rangle_{P_0}-\left\langle v_i^{\prime} h_j^{\prime}\right\rangle_{P_0}
    \label{56}
\end{equation}
where the second term is
\begin{equation}
\begin{aligned}
      \left\langle v_i^{\prime} h_j^{\prime}\right\rangle_{P_0}=\sum_{v, h, v^{\prime}, h^{\prime}} v_i^{\prime} h_j^{\prime} P\left(h^{\prime} \mid v^{\prime}\right) P\left(v^{\prime} \mid h\right) P(h \mid v) P_0(v)\\
      \approx \frac{1}{N}\sum_{v, h, v^{\prime}, h^{\prime}} v_i^{\prime} h_j^{\prime} P\left(h^{\prime} \mid v^{\prime}\right) P\left(v^{\prime} \mid h\right) P(h \mid v) P_0(v)
\end{aligned}
\label{57}
\end{equation}
Similarly, the $v,h,v^{\prime},h^{\prime}$ appearing in the second equation of \eqref{57} refer to the state combinations that appear during Monte Carlo sampling. The approximation methods in \eqref{56} and \eqref{57} are also known as the \textbf{contrastive divergence} method. In summary, the weight update method during the training of an RBM is
\begin{equation}
    w_{ij}^{\prime}=w_{ij}-\eta \frac{\partial}{\partial w_{ij}}C
    \label{58}
\end{equation}
where $\eta$ is the learning rate.

Using this method, the training of an RBM becomes a process of first sampling the visible layer state $v$ using the empirical distribution $P_0(v)$, followed by a few steps of Monte Carlo sampling. This is another clever application of the idea of random sampling to solve the difficulty of exponential explosion. Of course, the cost function we defined here applies to the context of unsupervised learning with an input dataset. When solving for the ground state of quantum many-body systems, we need to use reinforcement learning. In the following chapters, we will use RBMs combined with reinforcement learning to compute the ground states of several types of spin models.

\section{Neural Quantum State in Variational Monte Carlo}
\label{sec3}
\subsection{Supervised \& Unsupervised Learning, and Reinforcement Learning}
\label{subsec3.1}
Before introducing how to use neural quantum states as ansatz in variational Monte Carlo (VMC), we need to clarify the modes of machine learning: \textbf{supervised learning, unsupervised learning, and reinforcement learning}, which will be used later in the VMC method to find the ground state. Unsupervised learning involves data without explicit "correct answers," aiming to discover structures, patterns, and relationships within the data. Supervised learning relies on labeled training data, meaning each input data has a corresponding output label (or "correct answer"), and the model's task is to learn the mapping between these inputs and outputs. In numerical calculations of quantum many-body systems, especially for large or strongly correlated systems, we usually cannot know the desired ground state in advance, so in the following calculation examples, we adopt the mode of unsupervised learning. In the context of neural quantum states, we can clearly distinguish the two learning modes by illustrating the corresponding cost functions in quantum state tomography for both cases.

In supervised learning, the target wave function $\left|\Psi_{\operatorname{tar}}\right\rangle$ is given in advance. Using $\left|\Psi_{\mathrm{NN}}(\pmb{\alpha})\right\rangle$ to express the variational wave function represented by the neural network depending on a series of parameters $\pmb{\alpha}$, the cost function is defined as \cite{Carleo_2019}
\begin{equation}
    \mathcal{L}(\pmb{\alpha})=-\log\left[\frac{\left\langle\Psi_{\mathrm{tar}} \mid \Psi_{\mathrm{NN}}(\pmb{\alpha})\right\rangle}{\left\langle\Psi_{\mathrm{tar}} \mid \Psi_{\mathrm{tar}}\right\rangle} \frac{\left\langle\Psi_{\mathrm{NN}}(\pmb{\alpha}) \mid \Psi_{\mathrm{tar}}\right\rangle}{\left\langle\Psi_{\mathrm{NN}}(\pmb{\alpha}) \mid \Psi_{\mathrm{NN}}(\pmb{\alpha})\right\rangle}\right]
\end{equation}
The ultimate goal of training is to minimize this cost function value as much as possible. Its calculation is completed through the Monte Carlo method: sampling based on the probability amplitude of each configuration $|x\rangle$ in the target wave function and finally calculating the average result obtained from several samples. We can then optimize the parameters using its gradient, for example, in stochastic gradient descent
\begin{equation}
    \pmb{\alpha} \rightarrow \pmb{\alpha}-\lambda \pmb{\nabla}_{\pmb{\alpha}} \mathcal{L}
\end{equation}
where $\lambda$ is the learning rate. As seen, in supervised learning, the cost function makes the neural network imitate known rules.

In unsupervised learning, we do not pre-set a target wave function, meaning we do not know the probability of each configuration $|x\rangle$ in advance, which is somewhat similar to the scenario in quantum state tomography \cite{PhysRevA.64.052312}. Quantum state tomography attempts to completely reconstruct an unknown quantum state through a series of simple measurement results, but it requires a lot of computational resources, making it applicable only to smaller systems. Neural quantum states, due to their ability to represent high-dimensional data in a low-dimensional manner, are applied in quantum state tomography and have been proven to reconstruct quantum states with many qubits and high entanglement \cite{Torlai2018}. In this scenario, the cost function is defined as
\begin{equation}
    \mathcal{L}=\sum_{\pmb{\sigma}^b \in \mathcal{D}} \log \pi\left(\pmb{\sigma}^{\pmb{b}}\right)\quad \pi(\pmb{\sigma})=\frac{\left|\Psi_{\mathrm{NN}}(\pmb{\sigma})\right|^2}{\sum_{\pmb{\sigma}^{\prime}}\left|\Psi_{\mathrm{NN}}\left(\pmb{\sigma}^{\prime}\right)\right|^2}
\end{equation}
where $\pmb{D}$ is the given measurement data set, and the probability distribution $\pi(\pmb{\sigma})$ is provided by the neural quantum state. For this cost function, the part related to the data set is simple to calculate, while obtaining the probability distribution $\pi(\sigma)$ associated with the neural quantum state requires methods such as Markov chain Monte Carlo.

By comparing the cost function settings in quantum state tomography under known and unknown target quantum states, we can understand the logic of supervised or unsupervised learning for given measurement data. However, for solving the ground state of many-body systems, we neither know the physical laws in advance (understood here as the ground state wave function) nor have corresponding measurement data, so the logic of finding the ground state should be similar to reinforcement learning.

In VMC method to solve the ground state of quantum many-body systems, the goal is usually to minimize the energy expectation value corresponding to the Hamiltonian $\mathcal{H}$ of the system. In the context of reinforcement learning, we need to clarify its agent, environment, state, action, and reward. The general case of reinforcement learning can be summarized as the process where the agent learns to make optimal decisions to obtain maximum rewards through trial and error in an environment. In the context of solving the ground state of many-body systems:

- Agent: The algorithm responsible for adjusting and optimizing the wave function parameters.

- Environment: Composed of the quantum system's Hamiltonian $\mathcal{H}$ and the possible state space.

- State: The current wave function of the system $\Psi(\pmb{\alpha})$.

- Action: Adjusting the parameters $\pmb{\alpha}$ of the wave function.

- Reward: Usually the negative of the energy expectation value $(-\langle \mathcal{H}\rangle)$, meaning the agent's goal is to maximize the reward, thereby minimizing the energy expectation value.

Specifically, the calculation formula for the energy expectation value is
\begin{equation}
    \begin{aligned}
\langle\mathcal{H}\rangle & =\frac{\sum_{\pmb{\sigma}, \pmb{\sigma}^{\prime}} \Psi^*(\pmb{\sigma})\left\langle\pmb{\sigma}|\hat{H}| \pmb{\sigma}^{\prime}\right\rangle \Psi\left(\pmb{\sigma}^{\prime}\right)}{\sum_{\pmb{\sigma}}|\Psi(\pmb{\sigma})|^2} \\
& =\sum_{\pmb{\sigma}}\left(\sum_{\pmb{\sigma}^{\prime}}\left\langle\pmb{\sigma}|\mathcal{H}| \pmb{\sigma}^{\prime}\right\rangle \frac{\Psi\left(\pmb{\sigma}^{\prime}\right)}{\Psi(\pmb{\sigma})}\right) \frac{|\Psi(\pmb{\sigma})|^2}{\sum_{\pmb{\sigma}^{\prime}}\left|\Psi\left(\pmb{\sigma}^{\prime}\right)\right|^2} \\
& \approx\left\langle\sum_{\pmb{\sigma}^{\prime}}\left\langle\pmb{\sigma}|\mathcal{H}| \pmb{\sigma}^{\prime}\right\rangle \frac{\Psi\left(\pmb{\sigma}^{\prime}\right)}{\Psi(\pmb{\sigma})}\right\rangle_{\pmb{\sigma}}
\end{aligned}
\end{equation}
where $\pmb{\sigma}$ represents different configurations, and $\Psi(\pmb{\sigma})=\langle \pmb{\sigma}|\Psi\rangle$. The average in the last line over different configurations $\pmb{\sigma}$ refers to the average result of a Markov chain generated according to the distribution $\rho \propto |\langle \pmb{\sigma}|\Psi\rangle|^2$, as discussed in \ref{subsec1.2}. To achieve optimization, we need to take the Hamiltonian's expectation value as the cost function and calculate its gradient with respect to the parameters for optimization. For stochastic gradient descent, this process can be expressed as
\begin{equation}
    \pmb{\alpha} \rightarrow \pmb{\alpha}-\lambda \pmb{\nabla}_{\pmb{\alpha}} \mathcal{H}
\end{equation}
In our subsequent calculations, we will base our work on the NetKet software package \cite{10.21468/SciPostPhysCodeb.7,10.21468/SciPostPhysCodeb.7-r3.4}. It uses stochastic reconfiguration as a preconditioner, which can accelerate convergence and avoid local minima. Moreover, using multiple different stochastic reconfigurations can serve as an ensemble learning strategy, improving the stability and reliability of the solution by averaging the results of multiple runs.

\subsection{Spin Models and Restricted Boltzmann Machines}
\label{subsec3.2}
In \ref{subsec2.1}, we discussed the mathematical essence of many-body wave functions, which can be simply described as a mapping from a vector to a complex number. In practical calculations, we can view the many-body wave function as a black box: input a specific configuration $S$, and the wave function $\Psi(S)$ outputs an amplitude and a phase. This black box can certainly be a specific neural network. For spin-1/2 systems, where each lattice site's value is binary, a restricted Boltzmann machine (RBM) is a suitable neural network architecture.
\begin{figure}
    \centering
    {
        \includegraphics[width=0.6\textwidth]{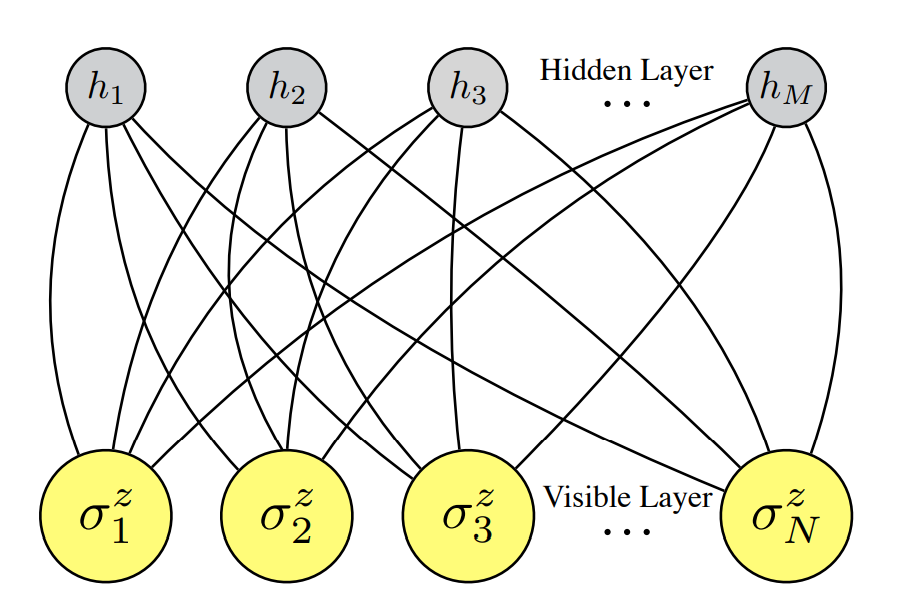}
    }
    \caption{The architecture of a restricted Boltzmann machine (RBM) \cite{doi:10.1126/science.aag2302}, consisting of a visible layer $\sigma^z_1,\sigma^z_2,...,\sigma^z_N$ representing the system's many-body quantum state and a hidden layer $h_1,h_2,...,h_M$. All neurons have binary values ($-1$ or $+1$) and there are no intra-layer interactions; neurons between layers are densely connected.}
    \label{RBM}
\end{figure}
The variational wave function form corresponding to the restricted Boltzmann machine is \cite{doi:10.1126/science.aag2302}
\begin{equation}
    \Psi_{RBM}(\mathcal{S} ; \mathcal{W})=\sum_{\left\{h_i\right\}}e^{-E(\mathcal{W})}=\sum_{\left\{h_i\right\}} e^{\sum_j a_j \sigma_j^z+\sum_i b_i h_i+\sum_{i j}h_i W_{i j}  \sigma_j^z}\quad h_i=\{-1,1\} 
    \label{59}
\end{equation}
where $\mathcal{W}$ represents the tunable parameters in the neural network, namely the weights $w_{ij}$ and biases $a_j, b_i$. In \ref{subsubsec2.2.4}, we discussed that the value of each neuron in an RBM can be either $1$ or $0$. Here, we use $\pm1$, which only affects the form of equations \eqref{53} and \eqref{54}. Using $\pm1$ makes the equations more symmetrical, and we will continue using $\pm1$ to represent spins up or down in the $z$ direction. Since there are no intra-layer neuron connections in the RBM, equation \eqref{59} can be directly written by separating the two cases of $h_i=\pm 1$ as
\begin{equation}
\begin{aligned}
   \Psi_{RBM}(\mathcal{S} ; \mathcal{W})=e^{\sum_j a_j \sigma^z_j} \times \prod_{i=1}^M F_i(\mathcal{S}) \\
   F_i(\mathcal{S})=2 \cosh \left[b_i+\sum_j W_{i j} \sigma_j^z\right]
\end{aligned}
\label{60}
\end{equation}
Note that the weights $w_{ij}$ in the network are defined as complex numbers to reasonably express both the amplitude and phase of the wave function.

This architecture has the following two main advantages \cite{doi:10.1126/science.aag2302}:
\begin{itemize}
    \item One practical advantage of this representation is that, in principle, the output quality of the neural network can be systematically improved by increasing the number of hidden variables. The number of units in the hidden layer $M$ (or equivalently, the density of hidden layer units $\alpha= M/N$) plays a role similar to the bond dimension in matrix product states. Additionally, the correlations induced by the hidden units are essentially highly non-local in real space (i.e., the lattice), making it suitable for describing quantum systems in any dimension.
    \item Another convenience of representing neural quantum states is that they can be expressed in a way that preserves symmetry \cite{sohn2012learning}. For example, the lattice's translational symmetry can be utilized to reduce the number of variational parameters in the neural quantum state ansatz, which follows the same idea as the translationally invariant RBM \cite{inproceedings}. Specifically, for integer hidden layer unit densities $\alpha = 1, 2, \ldots$, the weight matrix takes the form of convolution kernels $W^{(f)}_j$ discussed in \ref{subsubsec2.2.3}, where $f \in ([1, \alpha])$. These convolution kernels have a total of $\alpha N$ variational elements, rather than $\alpha N^2$ elements in the case without translational symmetry, as detailed in \ref{RBM Wavefunction Ansatz with Translation Symmetry}.
\end{itemize}

Since we generally cannot provide exact samples of the Hamiltonian's ground state, constructing a supervised learning (supervised learning) for the wave function $\Psi(s)$ is impractical. However, using a series of variational principles to find the ground state wave function can be achieved through reinforcement learning \cite{doi:10.1287/moor.22.1.222}. The wave function optimization method here is stochastic reconfiguration (SR) and stochastic gradient descent (SGD), with SR details in \ref{Reinforcement Learning Methods for Restricted Boltzmann Machines}. Each method has its advantages during optimization, so we can combine them. In practice, we usually choose SR as the preconditioner \cite{yang2015data} and SGD as the optimizer. SR can quickly approach the optimal solution in the initial optimization stage and is good at escaping local minima; SGD does not require the covariance matrix calculation, can quickly process large amounts of data through gradient calculation, and has some ability to escape saddle points \cite{dauphin2014identifying} due to its inherent noise model.
\subsubsection{Transverse-Field Ising Model}
\label{subsubsec3.2.1}
To verify the effectiveness of the RBM, we first use it to calculate two typical spin models - the transverse-field Ising model and the antiferromagnetic Heisenberg model, whose Hamiltonians are respectively
\begin{equation}
H_{\mathrm{TFI}}=-h \sum_i \sigma_i^x-J\sum_{\langle i, j\rangle} \sigma_i^z \sigma_j^z 
\label{61}
\end{equation}
where $\sigma^{l}, l=x, y, z$ are Pauli spin operators, and $\langle i, j\rangle$ represents a pair of nearest neighbor lattice sites.

For the transverse-field Ising model, we first calculate its ground state energy on a two-dimensional square lattice with periodic boundary conditions (PBC) and compare it with the results of exact diagonalization (ED). For convenience, we choose $J=1$ here.
\begin{figure}[H]
    \centering
    {
    \includegraphics[width=0.7\textwidth]{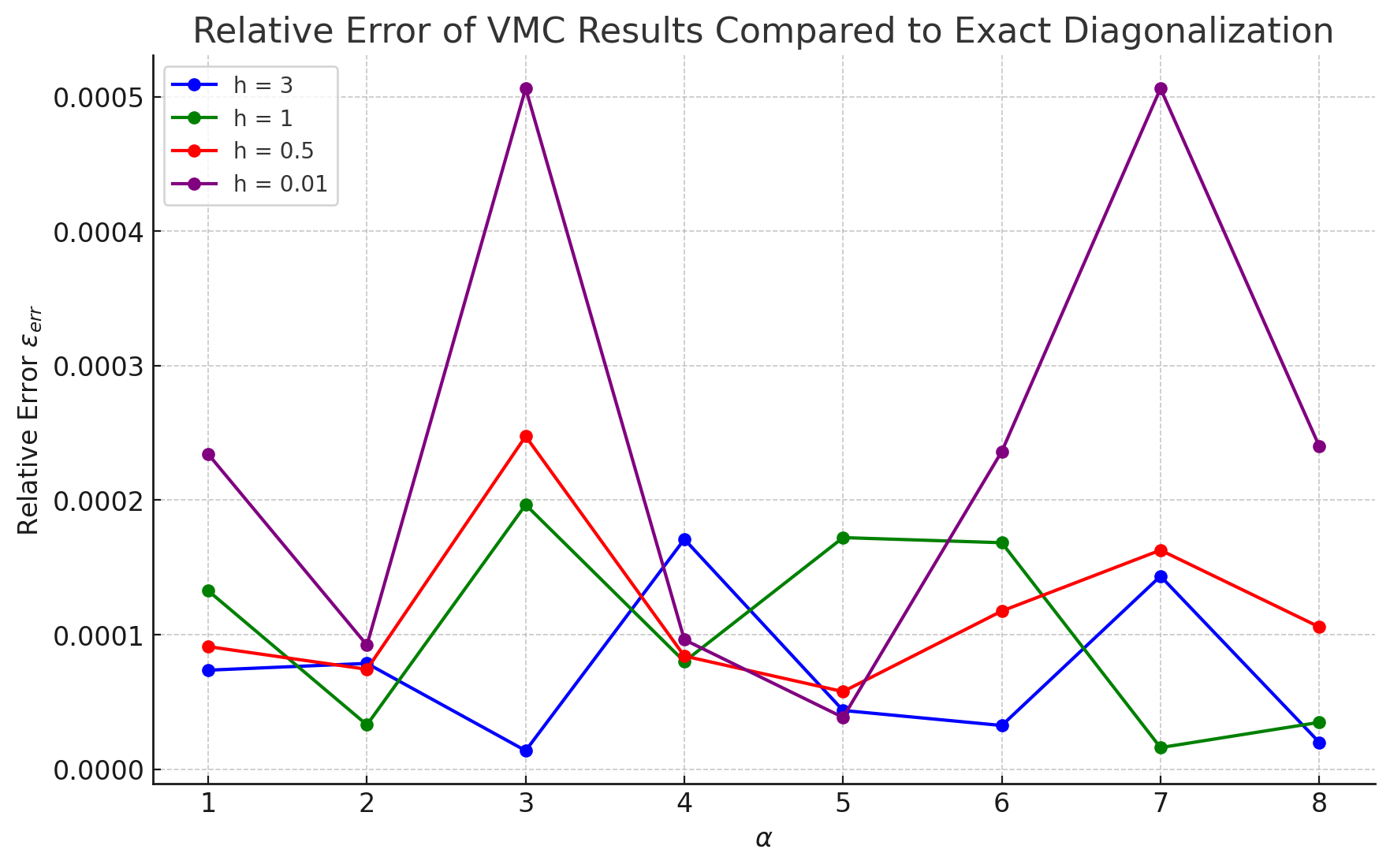}
    }
    \caption{Comparison of the ground state energy results of the transverse-field Ising model calculated using the RBM ansatz in the variational Monte Carlo method on a $L=5$ square lattice (PBC) with the results from exact diagonalization methods \cite{10.21468/SciPostPhys.2.1.003,10.21468/SciPostPhys.7.2.020}, reflected by the relative error $\varepsilon_{err}=|(E_{\text{VMC}}-E_{\text{ED}})/E_{\text{ED}}|$. The horizontal axis $\alpha$ represents the density of hidden layer units, and the vertical axis $h$ represents the magnetic field strength. All calculations were performed under the following parameter settings: preconditioner as stochastic reconfiguration with a diagonal offset of 0.001; optimizer using SGD with a linear schedule for learning rate adjustment, initial learning rate of 0.01, final learning rate of 0.0001, and learning period of 100; the first 10 results of each Markov chain were discarded. Specific ground state energy calculation results can be seen in \ref{Ground State Energy Calculations of the 2D Transverse Ising Model}}
    \label{gs_energy_of_2D_Transverse}
\end{figure}
As shown in \ref{gs_energy_of_2D_Transverse}, the RBM effectively calculates the ground state energy of the two-dimensional transverse-field Ising model with a small hidden layer unit density. However, the exact solution (quantum critical point) of the two-dimensional transverse-field Ising model is still somewhat controversial \cite{ZHANG2021114632}. To further confirm the accuracy of the RBM in finding the ground state, we can use this ansatz to calculate the quantum critical point of the one-dimensional transverse-field Ising model through the variational Monte Carlo method.

The quantum critical point of the one-dimensional transverse-field Ising model can also be obtained by mapping the self-dual property of the two-dimensional Ising model using the transfer matrix to the one-dimensional transverse-field Ising model. The zero-temperature ground state magnetically ordered and disordered boundary point is $h_c/J=1$ (see \ref{Analytical Solution of the 1D Transverse Ising Model} for details). Similarly, the two-dimensional transverse-field Ising model can be mapped to the three-dimensional Ising model, and this correspondence can be explained by finite-size scaling - the $d$-dimensional transverse-field Ising model and the $(d+1)$-dimensional Ising model belong to the same universality class (also connected through quantum-classical mapping \cite{Hsieh2012FromDQ}), sharing the same set of critical exponents, as detailed in \ref{A Brief Review of Phase Transitions and Scaling Laws}. According to the Mermin-Wagner theorem \cite{10.1063/1.1705316, PhysRevLett.17.1133}, spontaneous breaking of continuous symmetry in one-dimensional and two-dimensional systems is generally not possible at finite temperatures, meaning the system will not undergo a thermodynamic phase transition. However, the transverse-field Ising model has $Z_2$ symmetry, which is invariant under the transformation
\begin{equation}
    \sigma_j^x \rightarrow \mathcal{U}\sigma_j^x \mathcal{U}^{-1}=\sigma_j^x, \quad \sigma_j^z \rightarrow \mathcal{U}\sigma_j^z \mathcal{U}^{-1}=-\sigma_j^z
    \label{62}
\end{equation}
where the unitary transformation $\mathcal{U}$ is
\begin{equation}
    \mathcal{U}=\prod_{j}\sigma_j^x=\prod_{j}(-i) e^{i \pi S_j^x}
    \label{63}
\end{equation}
This symmetry is a discrete symmetry, meaning that for the transverse-field Ising model, thermodynamic phase transitions can exist, but not in the one-dimensional case \cite{Vojta_2003}. To understand the differences between one-dimensional and two-dimensional cases, we need to consider how domain walls in the system affect the stability of thermodynamic phase transitions. When considering macroscopic thermodynamic processes, we often need to consider the system's free energy $F$, which relates to internal energy $U$, thermodynamic temperature $T$, and entropy $S$ as
\begin{equation}
    F=U-TS
    \label{64}
\end{equation}
The direction of spontaneous changes in the system always reduces the free energy. Thus, we can estimate the change in free energy caused by the appearance of a domain wall excitation to roughly judge the preferred state of the spin system, which is the basis of Peierls' argument \cite{Peierls_1936, Bonati_2014}. According to this argument, in the one-dimensional case, the energy increase caused by adding a domain wall is much smaller than the corresponding huge increase in entropy. Therefore, the spin chain cannot resist the effects of thermal fluctuations at finite temperatures. In contrast, in two-dimensional and three-dimensional cases, the energy increase and entropy increase are comparable, allowing for the possibility of a magnetically ordered state at finite temperatures.
\begin{figure}[H]
    \centering
    {
    \includegraphics[width=0.75\textwidth]{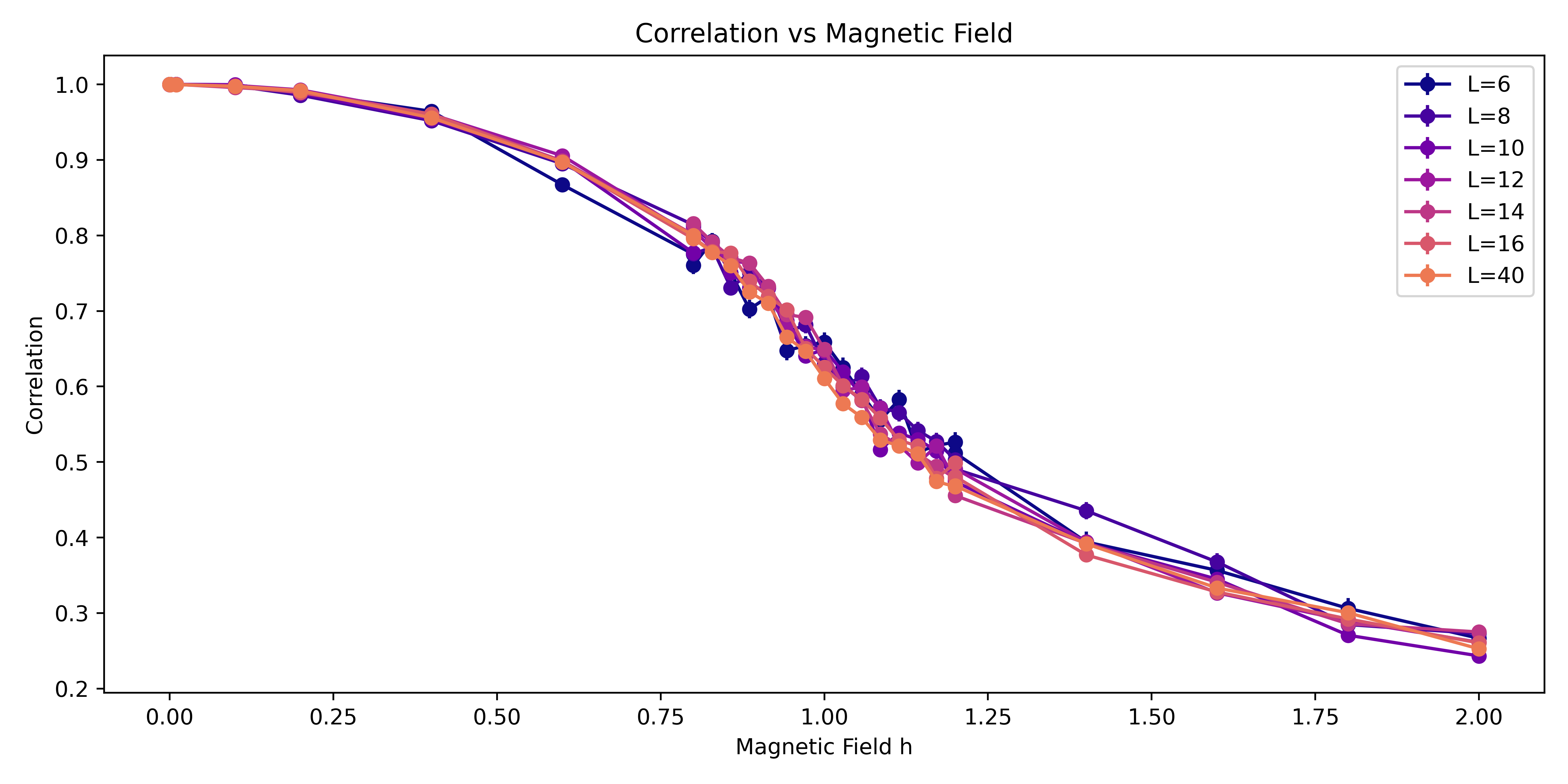}
    }
    \caption{The results of the nearest-neighbor correlation function $\langle \sigma^z_i\sigma^z_{i+1}\rangle$ calculated using the RBM ansatz in the variational Monte Carlo method for the transverse-field Ising model on one-dimensional chains of $L=6,8,10,12,14,16,40$ (PBC). It can be seen that there is a significant drop in the nearest-neighbor spin correlation near the quantum critical point $h_c/J=1$. All calculations were performed under the following parameter settings: preconditioner as stochastic reconfiguration with a diagonal offset of 0.001; optimizer using SGD with a linear schedule for learning rate adjustment, initial learning rate of 0.01, final learning rate of $10^{-6}$, and learning period of $10^4$; the first 3000 results of each Markov chain were discarded.}
    \label{NN_correlation}
\end{figure}
To preliminarily test the reliability of the neural quantum state variational Monte Carlo method, we first calculate the nearest-neighbor correlation function, which does not exhibit singular behavior near the critical point, to observe the response of the correlation function as the magnetic field strength approaches the critical point. As shown in \ref{NN_correlation}, when the magnetic field is near 0.001, 0.01, and 0.1, the ground state of the system is close to a completely ferromagnetic state. As the magnetic field increases, the quantum fluctuations introduced by the transverse field term $\sigma^x$ ($\sigma^x=\frac{1}{2}(\sigma^{+}+\sigma^{-})$) become more pronounced, leading to a significant decrease in the nearest-neighbor correlation function near the critical point.
\begin{figure}[H]
    \centering
    {
    \includegraphics[width=0.8\textwidth]{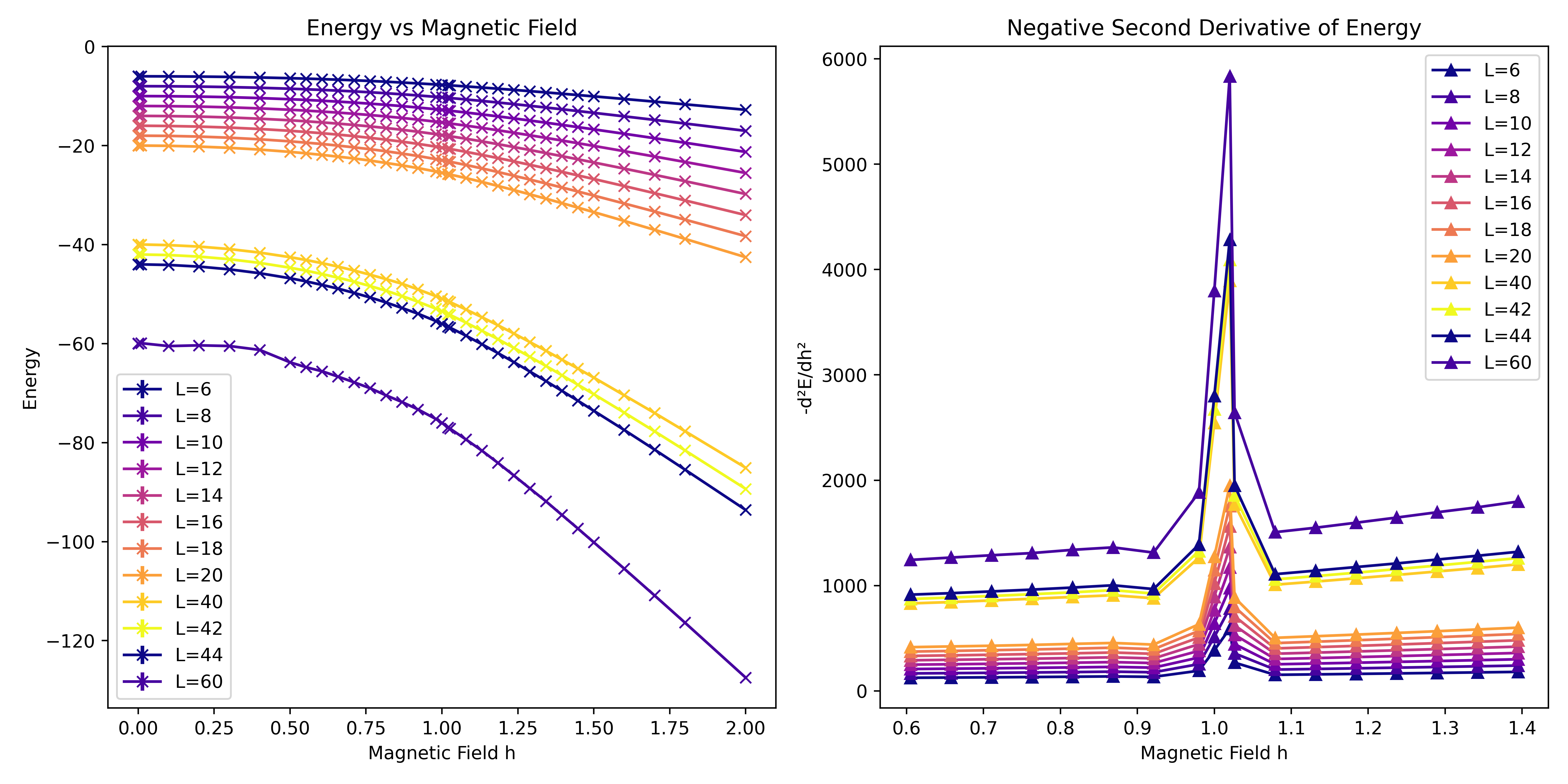}
    }
    \caption{The ground state energy $E$ and its negative second-order derivative with respect to the magnetic field $-\partial^2E/\partial h^2$ of the transverse-field Ising model on one-dimensional chains of $L=6,8,10,12,14,16,18,20,40,42,44,60$ (PBC). It can be seen that near $h=1.02$, the second-order derivative of the ground state energy tends to diverge in the thermodynamic limit, suggesting the existence of a second-order phase transition at this point. Machine learning and Markov chain Monte Carlo related parameter settings are the same as in Figure \ref{NN_correlation}.}
    \label{1D_Ising_energy}
\end{figure}
\begin{figure}[H]
    \centering
    {
    \includegraphics[width=0.8\textwidth]{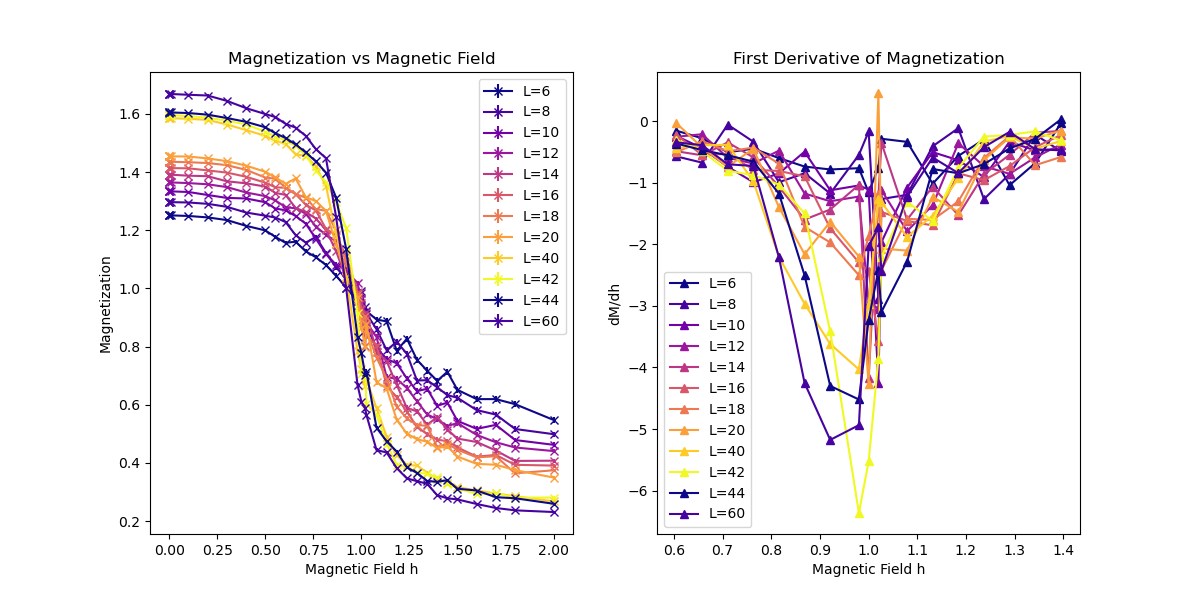}
    }
    \caption{The results of the scaled average magnetization $mL^{\beta/\nu}$ ($\beta=1/8, \nu=1, m=\frac{1}{L}\langle \sum_i \sigma^z_i \rangle$) and the magnetic susceptibility $\partial M/\partial h$ of the transverse-field Ising model on one-dimensional chains of $L=6,8,10,12,14,16,18,20,40,42,44,60$ (PBC). The parameter settings for machine learning and Markov chain Monte Carlo are the same as in Figure \ref{NN_correlation}. It is worth noting that in calculating the average magnetization, the absolute value of the expected value of the sum of $z$-direction spins in each spin configuration is averaged. As shown in the figure, the common intersection point of $mL^{\beta/\nu}$ for different scales is around $h=0.90$, which differs significantly from the exact theoretical value.}
    \label{1D_Ising_mag}
\end{figure}
To further examine the phase transition behavior near the quantum critical point $h_c=1$, we need to observe the behavior of the second-order derivative of the ground state energy with respect to the magnetic field $\frac{\partial^2E}{\partial h^2}$, the scaling of the average magnetization $mL^{\beta/\nu}$, and the magnetic susceptibility $\frac{\partial M}{\partial h}$ near the critical point. According to \ref{1D_Ising_energy} and \ref{1D_Ising_mag}, it can be seen that $\frac{\partial^2E}{\partial h^2}$ shows a significant tendency to diverge near $h=1.02$ (this quantity tends to infinity as the system approaches the thermodynamic limit $L\rightarrow \infty$). However, the tendency for the magnetic susceptibility to diverge does not always appear at a fixed point (the cases of $L=42, 44, 60$ in \ref{1D_Ising_mag} each give different peak points), and the common intersection point of the scaled average magnetization for different sizes is around $h=0.90$. Therefore, we can see that these two methods of determining the critical point cannot give a consistent critical point but only indicate the existence of a critical point.

\subsection{Spin Models and Convolutional Neural Networks}
\label{subsec3.3}
\subsubsection{Antiferromagnetic $J_1-J_2$ Model}
\label{subsubsec3.3.1}
Computing the ground state of frustrated magnetic systems has always been a concern, and the antiferromagnetic $J_1-J_2$ model is a typical spin model with a highly frustrated ground state. Its Hamiltonian is given by
\begin{equation}
    H=J_1 \sum_{\langle i j\rangle} \pmb{S}_i \cdot \pmb{S}_j + J_2 \sum_{\langle\langle i j\rangle\rangle} \pmb{S}_i \cdot \pmb{S}_j
    %\label{}
\end{equation}
where $\pmb{S}_i \cdot \pmb{S}_j = \sigma_i^x\sigma_j^x + \sigma_i^y\sigma_j^y + \sigma_i^z\sigma_j^z$, $\langle i,j\rangle$ represents a pair of nearest neighbor lattice sites, and $\langle \langle i,j\rangle \rangle$ represents a pair of next-nearest neighbor lattice sites. We are particularly interested in the case where $J_1, J_2 \textgreater 0$, i.e., both couplings are antiferromagnetic. In this case, when $J_1 \gg J_2$, the system does not exhibit frustration but rather long-range antiferromagnetic order \cite{PhysRevB.56.11678} (Néel order) with a pitch vector of $(\pi,\pi)$ (in two dimensions, this means spins are antiferromagnetically arranged in both the $x$ and $y$ directions). When $J_2 \gg J_1$, the system still shows magnetic order, with a striped order corresponding to a pitch vector of $(\pi,0)$ or $(0,\pi)$ \cite{Morita_2015}. However, in the intermediate case where $J_1$ and $J_2$ are comparable, i.e., around $J_2/J_1 \approx 0.5$, the ground state is highly frustrated. As of the writing of this article, several conflicting ground state proposals have been made for this situation: the plaquette valence-bond state \cite{PhysRevB.54.9007, PhysRevB.85.094407} (PVBS), the columnar valence-bond state \cite{PhysRevB.41.9323, PhysRevB.60.7278} (CVBS), and the gapless quantum spin liquid \cite{PhysRevLett.87.097201, PhysRevB.98.241109}. However, there is no clear answer yet.

For such highly frustrated systems, various ad hoc ansatz used in traditional VMC, or different types of projectors, such as the Gutzwiller projection \cite{PhysRevLett.10.159, PhysRev.137.A1726}, seem subjective. Neural networks, by virtue of their large number of parameters, can well compensate for the subjectivity of preconceived assumptions about a system we do not fully understand. In theory, a network with many parameters can achieve arbitrarily accurate solutions. However, to complete the optimization in a reasonable time, it is necessary to add some symmetry constraints to exclude invalid solutions. For lattice systems, this corresponds to lattice point symmetries.
\begin{figure}[H]
    \centering
    \subfloat[Schematic of the square lattice $J_1-J_2$ model]{
        \includegraphics[width=0.3\textwidth]{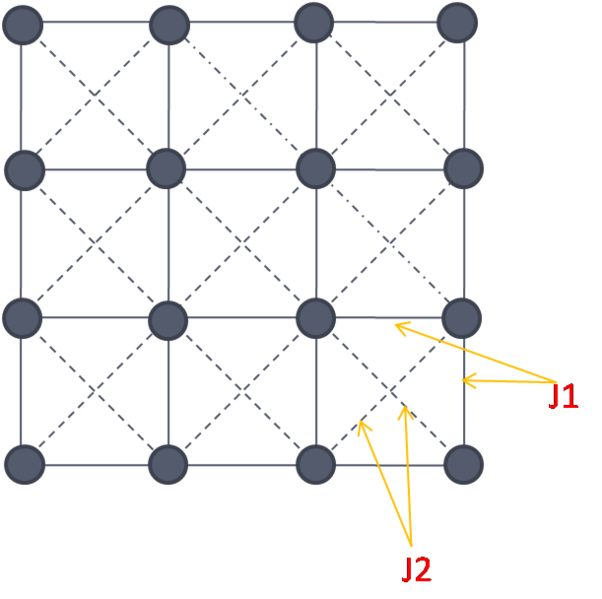}
    }
    \subfloat[Possible magnetic ordered states in the square lattice $J_1-J_2$ model]{
        \includegraphics[width=0.6\textwidth]{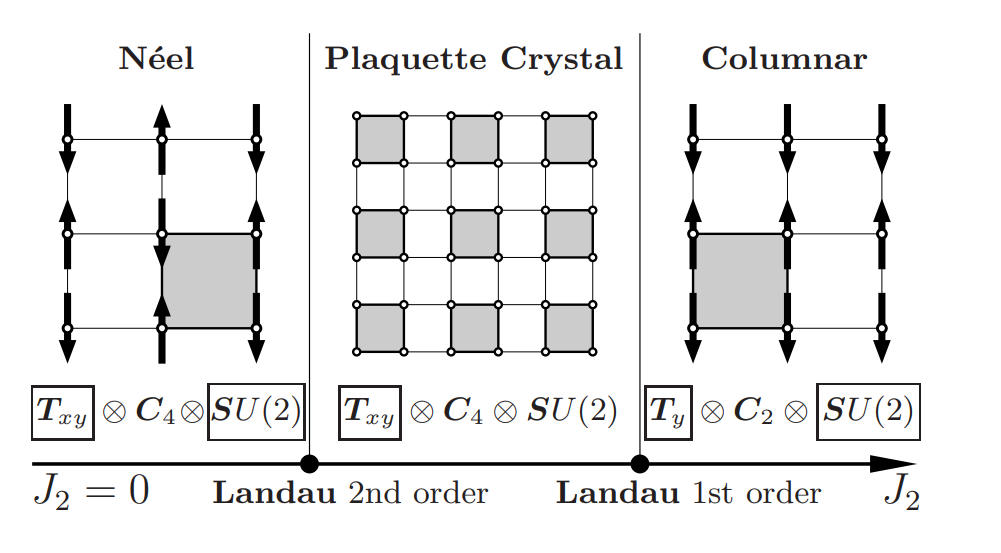}
    }
    \caption{Square lattice $J_1-J_2$ model and its possible phase diagrams: (a) The square lattice $J_1-J_2$ model has two types of links \cite{enwiki:1221074116}, namely nearest neighbor (N.N.) and next-nearest neighbor (N.N.N.); (b) Possible magnetic ordered states in the square lattice $J_1-J_2$ model \cite{article}, from left to right: Néel state, plaquette valence-bond state, and columnar valence-bond state. In the Néel state, each spin on the lattice is opposite to its nearest neighbor, breaking the translational symmetry in the $x$ and $y$ directions (spontaneous symmetry breaking, SSB), and $\pmb{SU}(2)$ symmetry (breaking the isotropy of the 1/2 spin breaks the $\pmb{SU}(2)$ symmetry). However, it does not break the fourfold rotational symmetry $\pmb{C}_4$ (rotating four times by $\pi/2$ around the center of the square yields the same spin configuration). In the plaquette valence-bond state, spins on four lattice sites form singlets pairwise, and a singlet pairing is called a valence bond, breaking only the translational symmetry in the $x$ and $y$ directions. For the columnar valence-bond state, singlets are formed in the $y$ (or $x$) direction, breaking only the translational symmetry in the $y$ (or $x$) direction and the $\pmb{SU}(2)$ symmetry. Increasing $J_2$ on the basis of forming a valence bond solid (VBS) (generally refers to all quantum spin liquid states forming valence bonds, breaking part or all of the translational symmetry but not the $\pmb{SU}(2)$ symmetry, including the three ground state proposals for $J_2/J_1 \approx 0.5$), a striped state as shown on the far right of the figure will emerge.}
    \label{gs_J1J2}
\end{figure}
The ground state and low-energy excitations of many-body systems often conform more to lattice symmetries. Some studies have shown \cite{Luo_2021,luo2023gauge,Vieijra_2020} that adding symmetry constraints to neural networks can greatly improve their ability to characterize system ground states, and even find low-energy excitations (achievable by penalizing orthogonality of trial wavefunctions with the ground state or directly training in different symmetry sectors). Group convolutional neural networks \cite{cohen2016group} (GCNNs) build on convolutional neural networks (CNNs) by implementing equivariant convolution, a complex linear operation that preserves various discrete group symmetries. In \ref{subsubsec2.2.3}, we have briefly understood the architecture of CNNs. Like all neural networks, their overall function can be expressed as a linear mapping $\Xi_{\mathrm{CNN}}:\{\pmb{\sigma}\} \rightarrow \mathbb{C}$, where $\pmb{\sigma}$ is the basis of the Hilbert space. Thus, we set the variational wavefunction as
\begin{equation}
    \Psi_{\mathrm{CNN}}(\pmb{\sigma})=\exp \left[\Xi_{\mathrm{CNN}}(\pmb{\sigma})\right]
\end{equation}
On this basis, we add considerations of lattice point group symmetries to construct group convolutional networks. CNNs have equivariance under translational groups, where the value of the neurons is determined by the convolution kernel and the values of the neurons in the receptive field in the previous layer \cite{roth2021group}
\begin{equation}
    C_{x, y}^i = \sum_{x^{\prime}, y^{\prime}<L} \pmb{W}_{x^{\prime}-x, y^{\prime}-y}^i \cdot \pmb{f}_{x^{\prime}, y^{\prime}}
\end{equation}
where the receptive field is required to satisfy periodic boundary conditions when sweeping through the neurons in the previous layer, corresponding to our physical system requirements. Here, $i$ represents different channels, $\pmb{f}$ represents the feature map of the previous layer, $C_{x,y}^i$ represents the feature map of the next layer with channel distinction, and $x^{\prime},y^{\prime}$ indicate the receptive field. Now we need to introduce more complex point group symmetries beyond translational symmetry, which may be non-Abelian and may include mirror symmetry operations and rotational operations. Let $g$ denote an element of such a discrete group $G$, then the group convolution operation can be written as
\begin{equation}
    C_g^i = \sum_{h \in G} \pmb{W}_{g^{-1} h}^i \cdot \pmb{f}_h
\end{equation}
Thus, the feature maps here are defined on the group space $G$ generated by the wallpaper group, not just on the translational group. By performing a group transformation $u$ on the convolution kernel
\begin{equation}
    \sum_{h \in G} \pmb{W}_{g^{-1} h}^i \cdot \pmb{f}_{u h} = \sum_{h \in G} \pmb{W}_{g^{-1} u^{-1} h}^i \cdot \pmb{f}_h = C_{u g}^i
\end{equation}
we can see that the transformation applied to the input feature map $\pmb{f}$ is also applied to the output feature map $C$, indicating that group convolution is also an equivariant operation.

Now we need to use GCNN to make a new variational wavefunction ansatz. Using $\Gamma$ to represent the non-linear activation function on each neuron, the first group convolution from the system configuration $\pmb{\sigma}$ can be expressed as
\begin{equation}
    \pmb{f}_g^1 = \Gamma\left(\sum_{\pmb{x}} W_{g^{-1} \pmb{x}}^0 \sigma_{\pmb{x}}\right)
\end{equation}
where the superscript of the feature map and the convolution kernel represents the corresponding layer number, and $\pmb{x}$ represents the lattice site position. It can be seen that there is a corresponding convolution kernel for each symmetry operation of the wallpaper group, thus achieving the group convolution operation from the input to the first feature map. Then, we need to perform group convolution from feature map to feature map
\begin{equation}
    \pmb{f}_g^{i+1} = \Gamma\left(\sum_{h \in G} W_{g^{-1} h}^i \pmb{f}_h^i\right)
\end{equation}
repeating this process until the final $N$th layer, where the feature map $f_g^N$ corresponding to the group element $g$ has only one neuron, representing a complex number. Thus, our wavefunction ansatz can be written as
\begin{equation}
    \Psi_{\mathrm{GCNN}}(\pmb{\sigma}) = \sum_g \chi_{g^{-1}} \exp \left(\pmb{f}_g^N\right)
\end{equation}
where $\chi_g$ represents the character corresponding to the group element. In the square lattice with $C_4$ symmetry that we will calculate, it can be written as \cite{PhysRevB.100.125124}
\begin{equation}
    \Psi_{\mathrm{GCNN}}(\pmb{\sigma}) = \sum_{r=0}^3 \chi_{c_4}^r \Psi_{\mathrm{CNN}}(\hat{c}^r_4 \pmb{\sigma})
\end{equation}
Here, the $C_4$ group is an Abelian group, with $\chi_{c_4} = e^{i\pi/2}$. In the subsequent calculations, we will choose the activation function $\Gamma$ as the SELU function \cite{klambauer2017selfnormalizing} to handle the case where the input to the neurons is a complex number. It was originally proposed to keep the output of each layer of the neural network having the same expectation and variance during training, thus helping to solve the gradient vanishing and gradient explosion problems in deep neural networks. Its expression is
\begin{equation}
    \Gamma(x) = \operatorname{SELU}(\operatorname{Re}(x)) + i \operatorname{SELU}(\operatorname{Im}(x))
\end{equation}
where
\begin{equation}
    \operatorname{SELU}(x) = \lambda \begin{cases}x & \text{ if } x > 0 \\ \alpha e^x - \alpha & \text{ if } x \leqslant 0\end{cases}
\end{equation}
The parameter values recommended by the original paper \cite{klambauer2017selfnormalizing} are $\lambda \approx 1.0507, \quad \alpha \approx 1.67326$ to maintain the self-normalizing property of the network.

For convenience, we set $J_1 = 1$, and use the structure factor
\begin{equation}
    S(\pmb{q}) = \frac{1}{N_{\pmb{s}}} \sum_{i, j} e^{i \pmb{q} \cdot \left(\pmb{r}_i - \pmb{r}_j\right)} \left\langle \pmb{S}_i \cdot \pmb{S}_j \right\rangle
\end{equation}
to determine the magnetic order of the ground state of the system for different $J_2$, where $\pmb{q}$ is called the pitch vector, and $N_s$ is the total number of lattice sites.

\begin{figure}[H]
    \centering
    \subfloat[Structure factor $J_2 = 0.2$]{
        \includegraphics[width=0.33\textwidth]{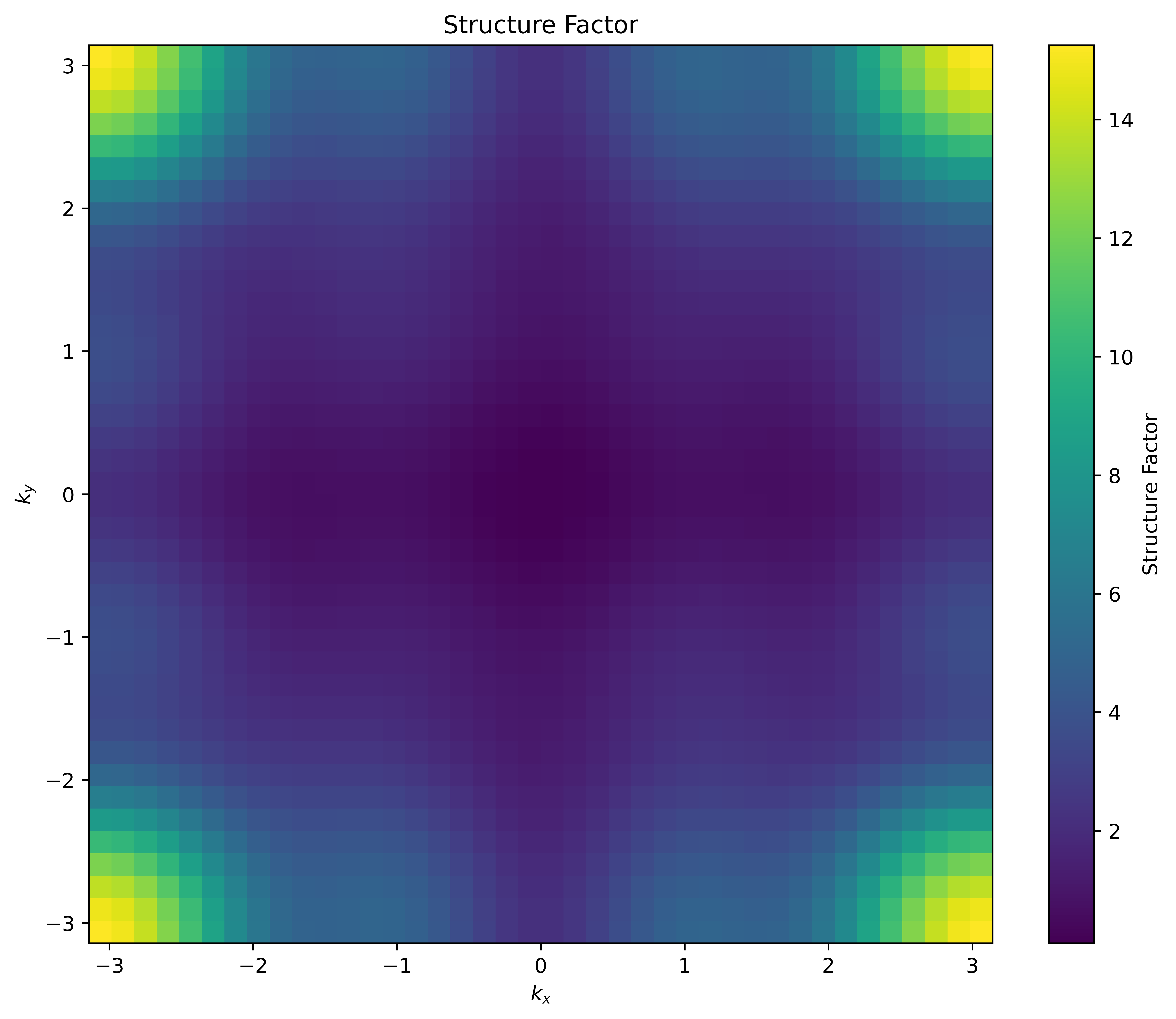}
    }
    \subfloat[Structure factor $J_2 = 0.8$]{
        \includegraphics[width=0.33\textwidth]{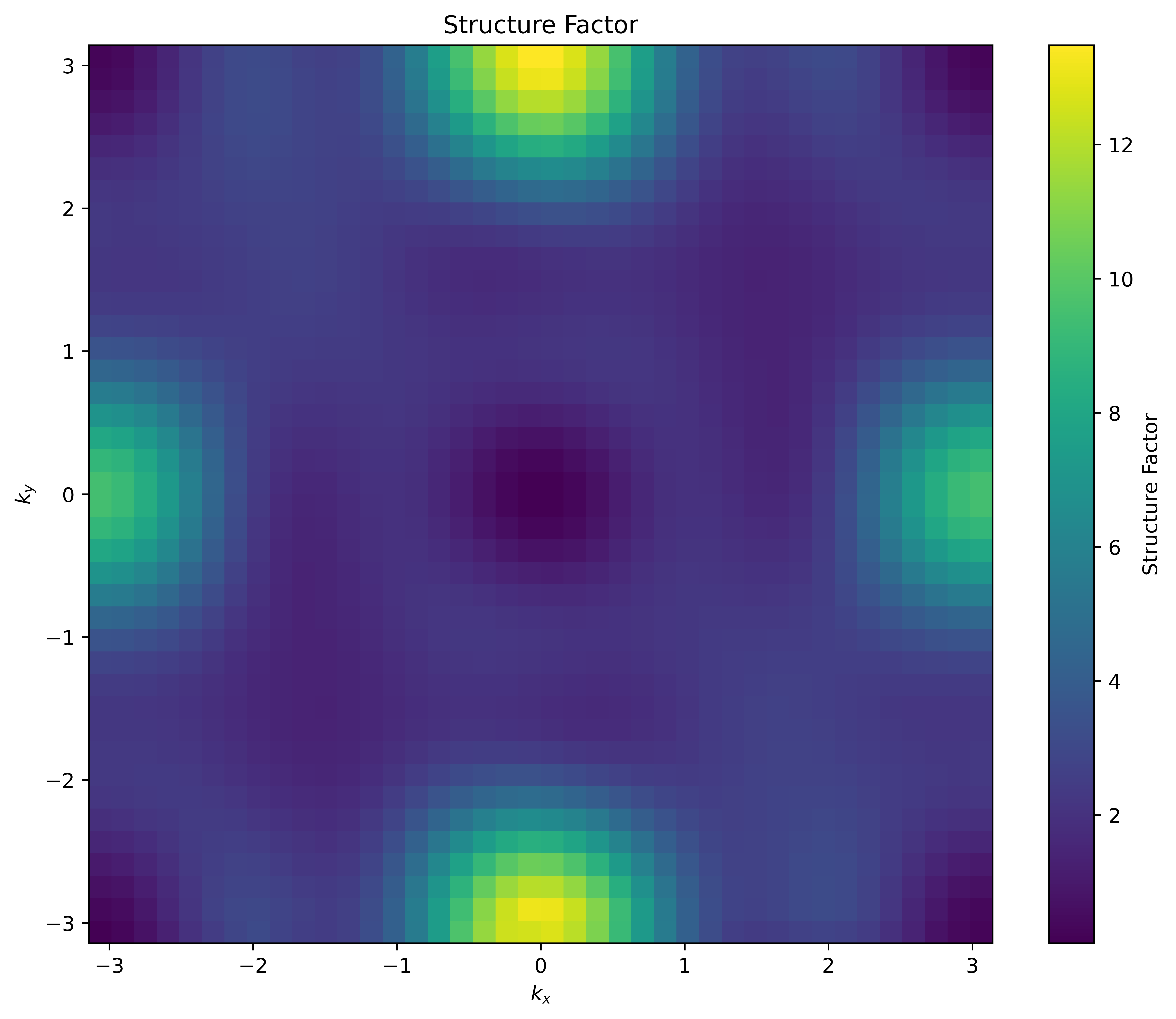}
    }
    \subfloat[Structure factor $J_2 = 0.5$]{
        \includegraphics[width=0.33\textwidth]{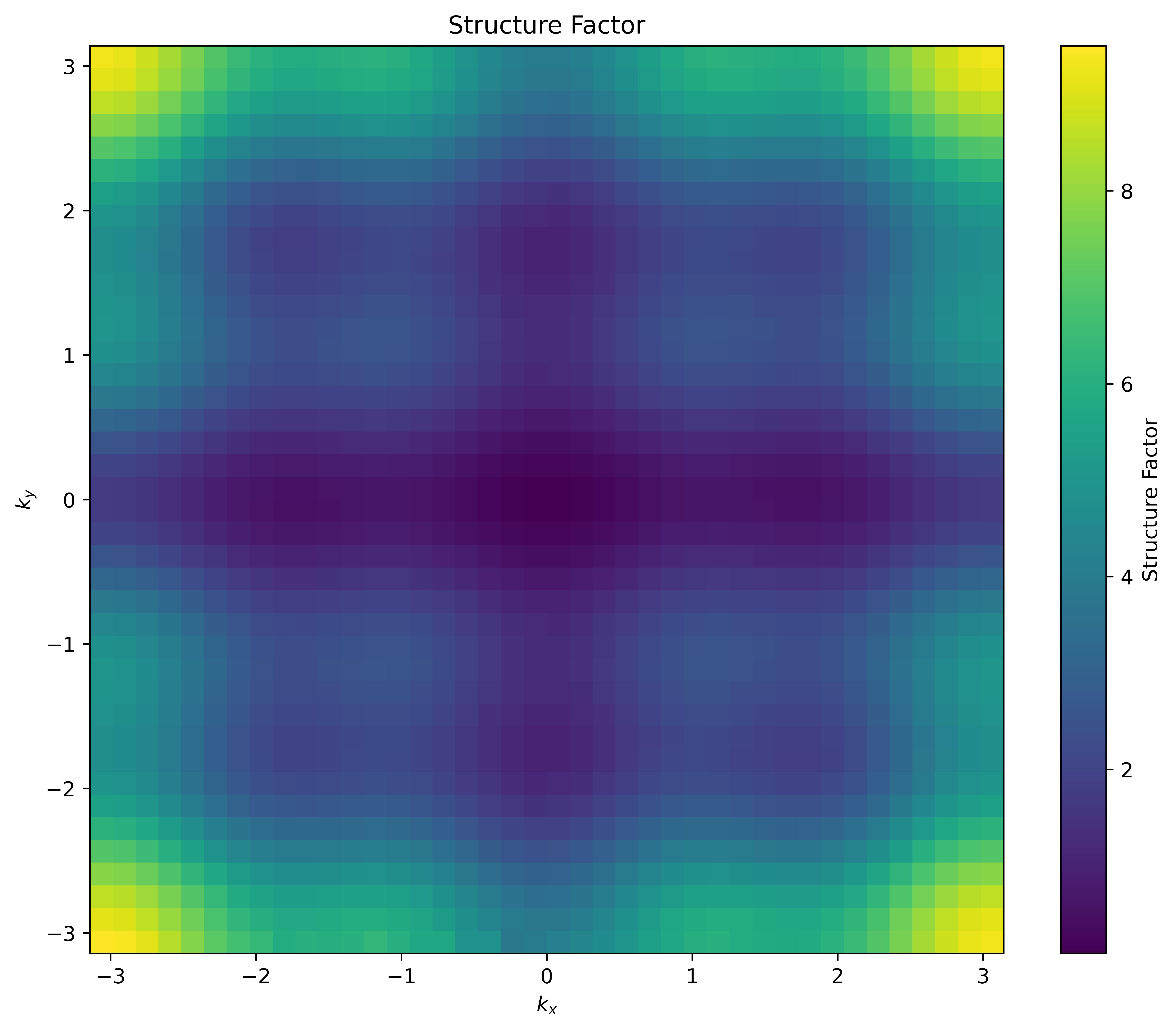}
    }
    \caption{Using GCNN combined with VMC to calculate the structure factor of the $J_1-J_2$ model on a square lattice with $L_x=L_y=4$ for different $J_2$: (a) $J_2=0.2$, the peak of the structure factor appears at $\pmb{q}=(\pm \pi,\pm \pi)$, indicating the presence of Néel antiferromagnetic order; (b) $J_2=0.8$, the peak of the structure factor appears at $\pmb{q}=(0,\pm \pi)$ or $\pmb{q}=(\pm \pi,0)$, indicating that the ground state is a striped state; (c) $J_2=0.5$ near the critical point, the structure factor does not show very obvious peaks as in the previous two cases, but its peaks are symmetrical in the $x$ and $y$ directions, mainly concentrated at $\pmb{q}=(\pm \pi,\pm \pi)$, suggesting that the ground state is more similar to the plaquette valence-bond state.}
    \label{L=4J1J2}
\end{figure}

\begin{figure}[H]
    \centering
    \subfloat[Structure factor $J_2 = 0.2$]{
        \includegraphics[width=0.33\textwidth]{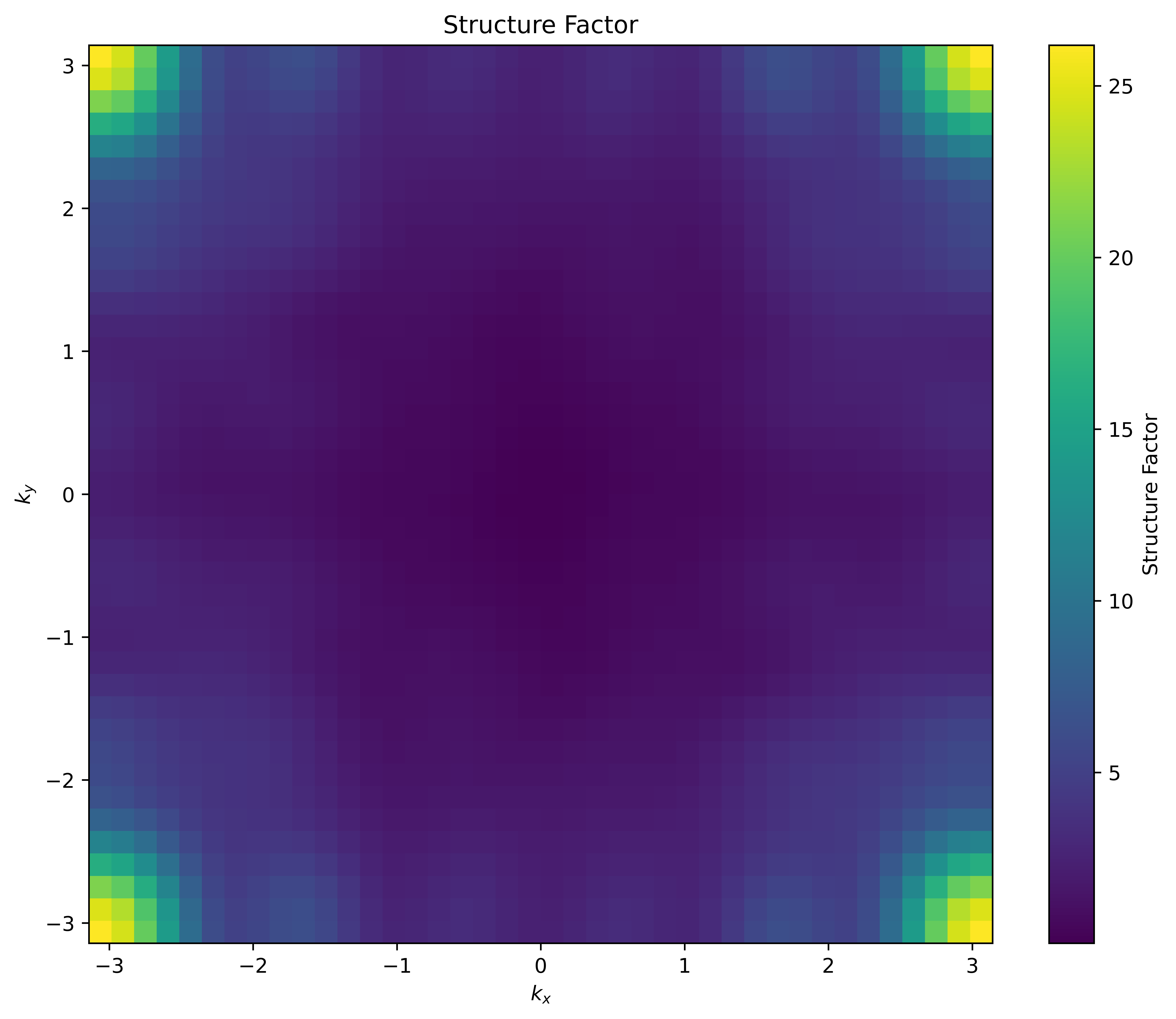}
    }
    \subfloat[Structure factor $J_2 = 0.8$]{
        \includegraphics[width=0.33\textwidth]{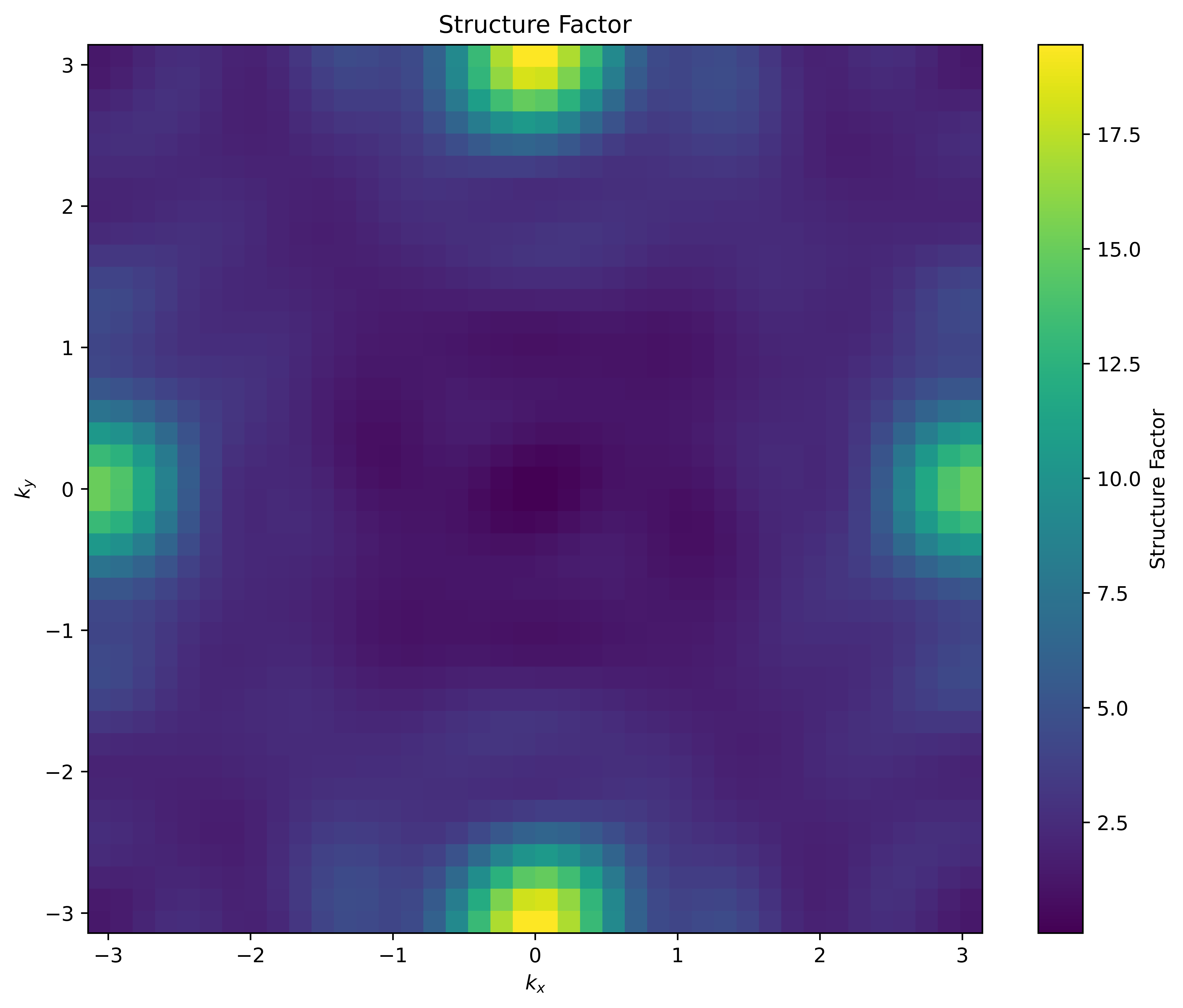}
    }
    \subfloat[Structure factor $J_2 = 0.5$]{
        \includegraphics[width=0.33\textwidth]{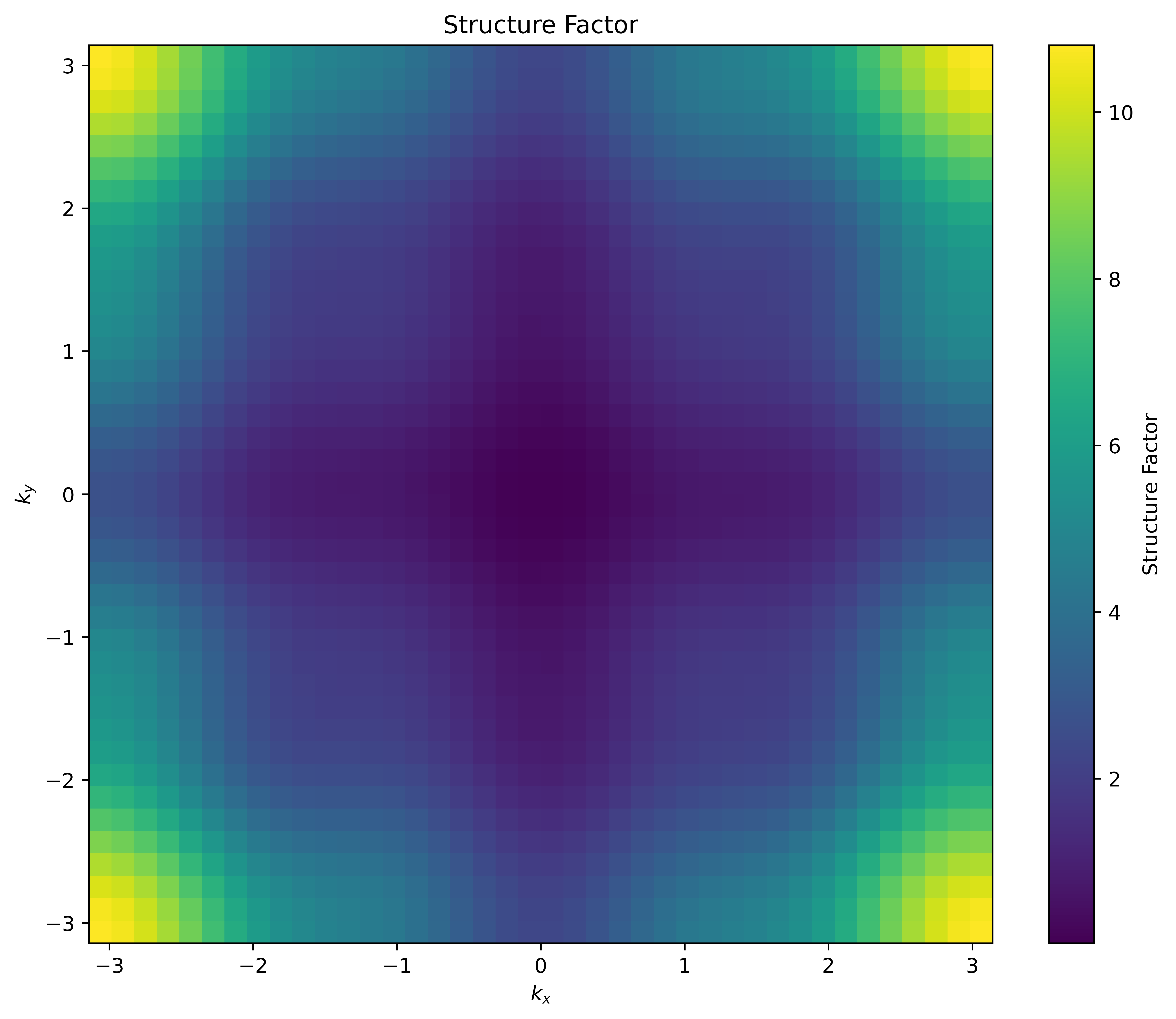}
    }
    \caption{Using GCNN combined with VMC to calculate the structure factor of the $J_1-J_2$ model on a square lattice with $L_x=L_y=6$ for different $J_2$: (a) $J_2=0.2$, the peak of the structure factor appears at $\pmb{q}=(\pm \pi,\pm \pi)$, indicating the presence of Néel antiferromagnetic order; (b) $J_2=0.8$, the peak of the structure factor appears at $\pmb{q}=(0,\pm \pi)$ or $\pmb{q}=(\pm \pi,0)$, indicating that the ground state is a striped state; (c) $J_2=0.5$ near the critical point, the structure factor does not show very obvious peaks as in the previous two cases, but its peaks are symmetrical in the $x$ and $y$ directions, mainly concentrated at $\pmb{q}=(\pm \pi,\pm \pi)$, suggesting that the ground state is more similar to the plaquette valence-bond state.}
    \label{L=6J1J2}
\end{figure}

\begin{figure}[H]
    \centering
    {
        \includegraphics[width=0.6\textwidth]{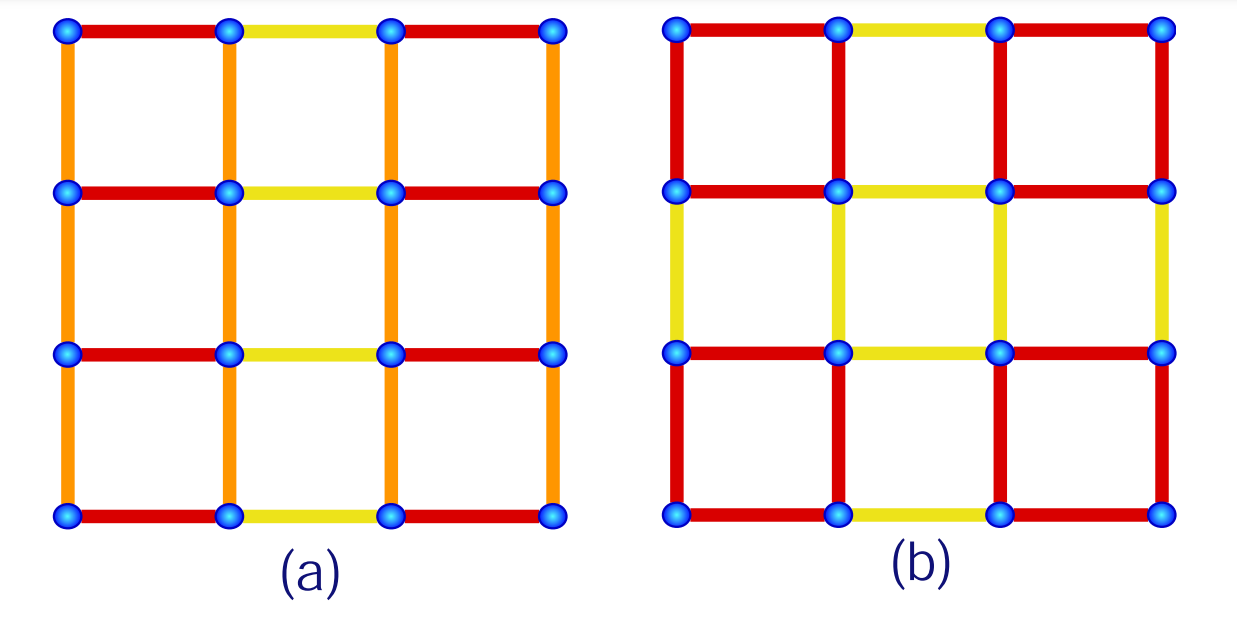}
    }
    \caption{Two possible VBS states \cite{doi:10.1126/science.1091806}: (a) Columnar valence-bond state, breaking $\pmb{SU}(2)$ symmetry but breaking translational symmetry in one direction; (b) Plaquette valence-bond state, breaking translational symmetry in both directions.}
    \label{VBS}
\end{figure}

As can be seen from \ref{L=4J1J2}, \ref{L=6J1J2}, the structure factor clearly reflects the magnetic order of the ground state of the system away from the region around $J_2/J_1 \approx 0.5$. For the case with weak next-nearest neighbor coupling, the ground state tends to be the Néel state; for the case with strong next-nearest neighbor coupling, the ground state tends to be the columnar valence-bond state. However, to determine which specific VBS state the ground state is in, we need to calculate the dimer-dimer structure factor \cite{Jiang_2012}
\begin{equation}
    M_d^{\alpha \beta}(\pmb{k}, L) = \frac{1}{L^2} \sum_{i j} e^{i \pmb{k} \cdot \left(\pmb{r}_i - \pmb{r}_j\right)} \left( \left\langle B_i^\alpha B_j^\beta \right\rangle - \left\langle B_i^\alpha \right\rangle \left\langle B_j^\beta \right\rangle \right)
\end{equation}
where $\alpha, \beta$ represent the unit vectors in the $x$ or $y$ direction, $B_i^\alpha \equiv \pmb{S}_i \cdot \pmb{S}_{i+\alpha}$. VBS patterns are generally expected to be found at the pitch vectors $\pmb{k}_x = (\pi, 0)$ or $\pmb{k}_y = (0, \pi)$, so the above equation can be simplified to
\begin{equation}
    m_{d, a}^2(L) = \frac{1}{L^2} M_d^{a a} \left( \pmb{k}_a, L \right) \quad \pmb{k} = \pmb{k}_a, a = x, y
\end{equation}
Based on the above dimer-dimer structure factor, T. Senthil, Ashvin Vishwanath, Leon Balents, Subir Sachdev, and Matthew P. A. Fisher \cite{doi:10.1126/science.1091806} proposed a complex order parameter $m_{d, x} + i m_{d, y}$ to distinguish between the columnar valence-bond state and the plaquette valence-bond state (as shown in \ref{VBS}). In simple terms, if the real or imaginary part of this order parameter is particularly prominent, it means that the arrangement of the dimers is highly correlated along one direction, suggesting that the ground state tends to be a columnar valence-bond state. When both the real and imaginary parts are evenly distributed, the ground state tends to be a plaquette valence-bond state.
Using the neural quantum state VMC method, we calculated the structure factors of the $J_1-J_2$ model at different $J_2$ for $L=4$ and $L=6$. The results show that for $J_2=0.2$, the ground state of the system is essentially the same as the ground state of the antiferromagnetic Heisenberg model; for $J_2=0.8$, the ground state is a striped state. However, for $J_2=0.5$, in the transition region from Néel state to VBS state (or from VBS state to striped state), it is difficult to determine the specific form of the ground state due to the less clear peaks in the structure factor compared to the previous two cases.
Due to limited computational resources, the author cannot construct larger lattice systems. For the training of neural networks, whether supervised, unsupervised, or reinforcement learning, the use of GPU acceleration is very necessary, which is very apparent when training convolutional neural networks, including group convolutional networks.

However, even large-scale simulations cannot explain the form of the ground state in the region around $J_2 \approx J_1$ so far. Strong quantum fluctuations destroy the possibility of long-range order in the system, and it is challenging to make reasonable predictions about the ground state wavefunction using perturbative methods in this region. The sign problem also poses a significant challenge to quantum Monte Carlo methods. To date, exploring the complete phase diagram of the $J_1-J_2$ model remains an active research area.

\section{Further}
\label{sec4}
Overall, the construction of neural quantum state variational Monte Carlo can be summarized as follows:

1. Select an appropriate neural network and formulate the corresponding ansatz $\Psi(\mathbf{\sigma})$. For spin systems, this could be a restricted Boltzmann machine (or its symmetric version) or a convolutional neural network (or a group convolutional network), etc.

2. Use the neural network as the neural quantum state, i.e., the variational wavefunction $|\Psi(\alpha)\rangle$, and initialize its parameters reasonably. A common way to do this is using a normal distribution. Use this initialized wavefunction as the initial variational wavefunction to generate its corresponding Markov chain.

3. Use the stochastic reconfiguration method to simulate imaginary time evolution, changing the variational wavefunction to lower the energy expectation value. This step acts as a preconditioner to help the subsequent neural network optimizer (such as Adam, SGD) escape local minima and quickly converge.

4. Optimize the parameters based on the Hamiltonian expectation values calculated from the Markov chain. For example, when using the SGD optimizer, this process can be described as
\begin{equation}
    \pmb{\alpha} \rightarrow \pmb{\alpha}-\lambda \pmb{\nabla}_{\pmb{\alpha}} \mathcal{H}
\end{equation}
where the learning rate $\lambda$ can have different selection strategies. One strategy is to let it decrease gradually during multiple optimizations, which is often the method used in this paper.

The above process sounds straightforward, but in practice, the selection of various parameters requires considerable experience. The results in this paper correctly reflect some ground state characteristics of strongly correlated systems, but compared to other numerical methods, particularly some other quantum Monte Carlo methods, they seem lacking in precision and stability.

However, the low memory consumption characteristic of neural networks compels us to consider their potential for addressing the computation problems of very large systems. Some large-scale numerical computations require an overall grasp of the system's phase diagram rather than higher precision critical exponents. Perhaps in the future, when GPU performance improves further, this method can also be effective on a larger scale.

We also need to note that there are still many features that cannot be realized or are inconvenient to realize in the construction of the neural quantum state variational Monte Carlo method described in this paper. The most important one is calculating finite temperature systems. Fortunately, this problem has been solved using the combination of imaginary time evolution and time-dependent variational Monte Carlo (t-VMC) \cite{Takai_2016}, but its compatibility with neural quantum states remains to be evaluated. Another challenge is calculating the low excitation spectrum of the system. The author envisions that this might be addressed by introducing an orthogonality penalty to the previously found lower energy quantum state (by adding the inner product of the variational wavefunction with the solved ground state wavefunction to the cost function, to avoid the variational wavefunction approaching the ground state again, thereby obtaining the first excited state, and repeating the process). This approach has some obvious drawbacks: the penalty weight is difficult to determine, and it is challenging to handle systems where the low excited state and the ground state are very close. Of course, these problems require further exploration by subsequent researchers.

\bibliographystyle{elsarticle-num}
\bibliography{main}

\appendix

\section{}
\subsection{Minimum Value of Cross Entropy}
\label{Minimum Value of Cross Entropy}
When the predicted probability distribution \( q \) matches the true target probability distribution \( p \) exactly, i.e., each \( q_i = p_i \), it can be proven that the cross entropy \( S(p, q) \) reaches its minimum value. Specifically:

The definition of cross entropy is
\begin{equation}
 S(p, q) = -\sum_{i} p_i \log q_i   
\end{equation}

The definition of information entropy is
\begin{equation}
    S(p) = -\sum_{i} p_i \log p_i
\end{equation}

For all \( x > 0 \), the logarithm function satisfies Jensen's inequality
\begin{equation}
  \log x \leq x - 1  
\end{equation}

Replacing \( x \) with \( \frac{q_i}{p_i} \), we get:
\begin{equation}
    \frac{q_i}{p_i} \log \frac{q_i}{p_i} \geq \frac{q_i}{p_i} - 1
\end{equation}

Multiplying both sides by \( p_i \) and summing, we obtain:
\begin{equation}
  \sum_i q_i \log \frac{q_i}{p_i} \geq \sum_i (q_i - p_i)  
\end{equation}

Since \( \sum_i q_i = \sum_i p_i = 1 \), we have:
\begin{equation}
   \sum_i q_i \log \frac{q_i}{p_i} \geq 0 
   \label{A6}
\end{equation}

Rewriting \eqref{A6}:
\begin{equation}
   -\sum_i p_i \log q_i \leq -\sum_i p_i \log p_i 
\end{equation}

Therefore, the cross entropy \( S(p, q) \) reaches its minimum value \( S(p) \) when \( q_i = p_i \), i.e.:
\begin{equation}
  S(p, q) \geq S(p)  
\end{equation}

When \( q_i = p_i \), the above inequality becomes an equality because \( \frac{q_i}{p_i} = 1 \) makes \( \log \frac{q_i}{p_i} = 0 \). This shows that the cross entropy reaches its theoretical minimum when the predicted distribution is exactly the same as the true distribution. Due to this characteristic, cross entropy is used as the cost function when training restricted Boltzmann machines.

\subsection{Conditional Probability in Restricted Boltzmann Machines}
\label{Conditional Probability in Restricted Boltzmann Machines}
To derive equations \eqref{51} and \eqref{54}, a series of calculations is required, which we present here. In a Restricted Boltzmann Machine (RBM), given the visible layer state \( v \), the conditional probability of the hidden layer state \( h \) is derived from the joint probability distribution of the entire system. The calculation process for the conditional probability \( P(h \mid v) \) is as follows:

First, the joint distribution of the RBM is given by
\begin{equation}
P(v, h) = \frac{1}{Z} e^{-E(v, h)}
\end{equation}
where \( Z \) is the partition function of the system, defined as
\begin{equation}
Z = \sum_{v, h} e^{-E(v, h)}
\end{equation}
According to the definition of conditional probability, we have
\begin{equation}
P(h \mid v) = \frac{P(v, h)}{P(v)} = \frac{\frac{1}{Z} e^{-E(v, h)}}{\sum_h \frac{1}{Z} e^{-E(v, h)}}
\end{equation}
Simplifying, we get
\begin{equation}
P(h \mid v) = \frac{e^{-E(v, h)}}{\sum_h e^{-E(v, h)}}
\end{equation}
Here, the denominator is the sum of \( e^{-E(v, h)} \) over all possible hidden layer states \( h \) given \( v \), ensuring that the total conditional probability sums to 1.

For each hidden unit \( h_j \), we can write:
\begin{equation}
E(v, h)=-\sum_i a_i v_i-\sum_j b_j h_j-\sum_{i, j} v_i w_{i j} h_j
\end{equation}
In an RBM, each unit's value is either \( 0 \) or \( 1 \). Since the probability distribution of each hidden layer unit \( h_j \) is independent, we first calculate the probability distribution for a unit being \( h_j=1 \):
\begin{equation}
P(h_j = 1 \mid v) =\frac{e^{-E\left(v, h_j=1\right)}}{e^{-E\left(v, h_j=1\right)}+e^{-E\left(v, h_j=0\right)}}= \frac{e^{b_j + \sum_i v_i w_{ij}}}{1 + e^{b_j + \sum_i v_i w_{ij}}}
\end{equation}
And for \( h_j=0 \):
\begin{equation}
    P(h_j = 0 \mid v) =\frac{e^{-E\left(v, h_j=0\right)}}{e^{-E\left(v, h_j=1\right)}+e^{-E\left(v, h_j=0\right)}}= \frac{1}{1 + e^{b_j + \sum_i v_i w_{ij}}}
\end{equation}
Here we use \( z_j = b_j + \sum_i v_i w_{ij} \) to re-express:
\begin{equation}
P(h_j = 1 \mid v) = \frac{e^{z_j}}{1 + e^{z_j}},\quad P(h_j = 0 \mid v) = \frac{1}{1 + e^{z_j}}
\label{A16}
\end{equation}
\eqref{A16} can be summarized as
\begin{equation}
    P(h_j \mid v) = \frac{e^{z_j h_j}}{1 + e^{z_j}}
\end{equation}
Thus,
\begin{equation}
     P(h \mid v)=\prod_{j}P(h_j \mid v) =\prod_{j} \frac{e^{z_j h_j}}{1+e^{z_j}}
\end{equation}

\subsection{RBM Wavefunction Ansatz with Translation Symmetry}
\label{RBM Wavefunction Ansatz with Translation Symmetry}
Hamiltonians on lattices often exhibit inherent symmetries, which must be respected by their ground states and dynamically evolving quantum states. These symmetries can be utilized to reduce the number of variational parameters in neural quantum states.

We consider a symmetry group defined by a set of linear transformations \( T_s \), where \( s = 1, \ldots, S \), such that the spin configuration transforms according to \( T_s \sigma^z = \tilde{\sigma}^z(s) \). We can require the neural quantum state to remain invariant under the action of \( T_s \), defined as:
\begin{equation}
\Psi_\alpha(\mathcal{S} ; \mathcal{W}) = \sum_{\left\{h_{i, s}\right\}} \exp \left[\sum_f^\alpha a^{(s)} \sum_s^S \sum_j^N \tilde{\sigma}^z(s) + 
\sum_f^\alpha b^{(s)} \sum_s^S h_{f, s} + \sum_f \sum_S^S h_{f, s} \sum_j^N W_j^{(f)} \tilde{\sigma}^z(s) \right]
\end{equation}
where the weights of the neural network differ in dimension from those of a general neural quantum state. Specifically, \( a^{(f)} \) and \( b^{(f)} \) are vectors in the feature space, which is the space formed by the hidden units, with \( f = 1, \ldots, \alpha_s \). The convolutional kernel \( W_j^{(f)} \) contains \( \alpha_s \times N \) elements. Note that this expression is essentially equivalent to a general neural quantum state with \( M = S \times \alpha_s \) hidden variables. Summing over the hidden units, we obtain
\begin{equation}
\Psi_\alpha(\mathcal{S} ; \mathcal{W}) = e^{\sum_{f, s, j} a^{(f)} \widetilde{\sigma_j^z}(s)} \times \prod_f \prod_s 2 \cosh \left[b^{(f)} + \sum_j^N W_j^{(f)} \widetilde{\sigma_j^Z}(s)\right]
\end{equation}

In specific cases where there is lattice translation invariance, we have orbits of the symmetry group with \( S = N \) elements. For a given feature \( f \), the matrix \( W_j^{(f)} \) can be viewed as a convolutional kernel acting on the \( N \) translational replicas of a given spin configuration. In other words, each feature has a pool of \( N \) translation-related hidden units, which use the same convolutional kernel to act on the spin's symmetric transformations.

\subsection{Reinforcement Learning Methods for Restricted Boltzmann Machines}
\label{Reinforcement Learning Methods for Restricted Boltzmann Machines}
In \ref{subsec1.3}, we discussed how to find the ground state of the system using a method called 'stochastic reconfiguration'. For our variational wavefunction \( \ket{\Psi(\mathcal{S})} \), the expectation value \( E(\mathcal{W}) = \frac{\left\langle \Psi(\mathcal{S}) | \mathcal{H} | \Psi(\mathcal{S}) \right\rangle}{\left\langle \Psi(\mathcal{S}) | \Psi(\mathcal{S}) \right\rangle} \) is a function of the neural network weights \( \mathcal{W} \). To obtain the optimal solution that satisfies \( \nabla E(\mathcal{W}^{\star}) = 0 \), several optimization methods can be used.

Here, we find it convenient to adopt the stochastic reconfiguration (SR) method by Sorella et al. \cite{Mezzacapo_2009}, which can be interpreted as imaginary time evolution. Introducing the variational derivative with respect to the \( k \)-th network parameter,
\begin{equation}
\mathcal{O}_{\mathrm{k}}(\mathcal{\delta}) = \frac{1}{\langle \mathcal{S} | \Psi(\mathcal{S}) \rangle} \partial_{\mathcal{W}_{\mathrm{k}}} (\langle \mathcal{S} | \Psi(\mathcal{S}) \rangle)
\end{equation}
and the so-called local energy
\begin{equation}
\mathrm{E}_{\text {loc }}(\mathcal{S}) = \frac{\left\langle \mathcal{S} | \mathcal{H} | \Psi_{\mathrm{M}} \right\rangle}{\langle \mathcal{S} | \Psi(\mathcal{S}) \rangle}
\end{equation}
the SR update at the \( p \)-th iteration is given by
\begin{equation}
\mathcal{W}(p+1) = \mathcal{W}(p) - \gamma(p) S^{-1}(p) F(p)
\end{equation}
where we introduce a Hermitian covariance matrix
\begin{equation}
S_{k k^{\prime}}(p) = \left\langle \mathcal{O}_k^{\dagger} \mathcal{O}_{k^{\prime}} \right\rangle - \left\langle \mathcal{O}_k^{\dagger} \right\rangle \left\langle \mathcal{O}_{k^{\prime}} \right\rangle
\end{equation}
The expression for the forces is
\begin{equation}
F_k(p) = \left\langle E_{\text {loc }} \mathcal{O}_k^{\dagger} \right\rangle - \left\langle E_{\text {loc }} \right\rangle \left\langle \mathcal{O}_k^{\dagger} \right\rangle
\end{equation}
where \( \gamma(p) \) is a scaling parameter. Since the covariance matrix in equation (84) may be non-invertible, \( S^{-1} \) represents its Moore-Penrose pseudo-inverse. Additionally, a more precise regularization can be used, which takes the form \( S_{k, k^{\prime}}^{\text {reg }} = S_{k, k^{\prime}} + \lambda(p) \delta_{k, k^{\prime}} S_{k, k} \), where \( \lambda(p) \) corresponds to a step size factor that decays exponentially with the number of steps in the stochastic reconfiguration. The author prefers the latter method in practical work.

\section{}
\subsection{Ground State Energy Calculations of 2D Transverse Ising Model}
\label{Ground State Energy Calculations of the 2D Transverse Ising Model}
\begin{table}[H]
\centering
\setlength{\tabcolsep}{15pt} % Adjust the space between columns
\caption{Ground state energy values of the 2D transverse Ising model under different magnetic fields $h$ and hidden layer density $\alpha$}
\label{tab:energy_calculations}
\begin{tabular}{
  @{}
  l
  S[table-format=-2.4]
  S[table-format=-2.4]
  S[table-format=-2.4]
  S[table-format=-2.4]
  @{}
}
\toprule
L=5, Square Lattice & {h = 3} & {h = 1} & {h = 0.5} & {h = 0.01} \\
\midrule
Exact Diagonalization & -80.1331\quad & -53.1416\quad & -50.7823\quad & -50.0003 \\
\midrule
\(\alpha=1\) & -80.1390 \pm 0.0097 & -53.1486 \pm 0.0060 & -50.7869 \pm 0.0032 & -49.9886 \pm 0.0100 \\
\(\alpha=2\) & -80.1268 \pm 0.0084 & -53.1398 \pm 0.0046 & -50.7785 \pm 0.0055 & -49.9957 \pm 0.0068 \\
\(\alpha=3\) & -80.1320 \pm 0.0078 & -53.1520 \pm 0.0037 & -50.7697 \pm 0.0083 & -49.9750 \pm 0.0140 \\
\(\alpha=4\) & -80.1194 \pm 0.0071 & -53.1373 \pm 0.0069 & -50.7780 \pm 0.0059 & -49.9955 \pm 0.0068 \\
\(\alpha=5\) & -80.1366 \pm 0.0065 & -53.1324 \pm 0.0069 & -50.7852 \pm 0.0038 & -50.0022 \pm 0.0001 \\
\(\alpha=6\) & -80.1305 \pm 0.0068 & -53.1505 \pm 0.0058 & -50.7763 \pm 0.0051 & -49.9885 \pm 0.0098 \\
\(\alpha=7\) & -80.1216 \pm 0.0065 & -53.1407 \pm 0.0048 & -50.7740 \pm 0.0080 & -49.9750 \pm 0.0140 \\
\(\alpha=8\) & -80.1315 \pm 0.0068 & -53.1434 \pm 0.0063 & -50.7769 \pm 0.0075 & -49.9883 \pm 0.0098 \\
\bottomrule
\end{tabular}
\end{table}

\subsection{Analytical Solution of the 1D Transverse Ising Model}
\label{Analytical Solution of the 1D Transverse Ising Model}
\begin{figure}[H]
    \centering
    \subfloat[Dual Lattice]{\includegraphics[width=0.35\textwidth]{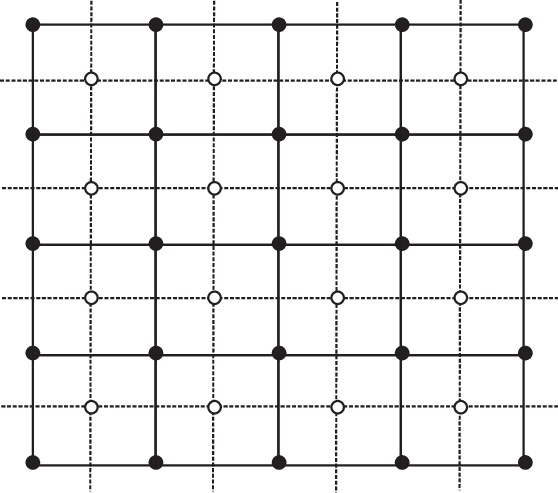}}
    \subfloat[Four Sites]{\includegraphics[width=0.35\textwidth]{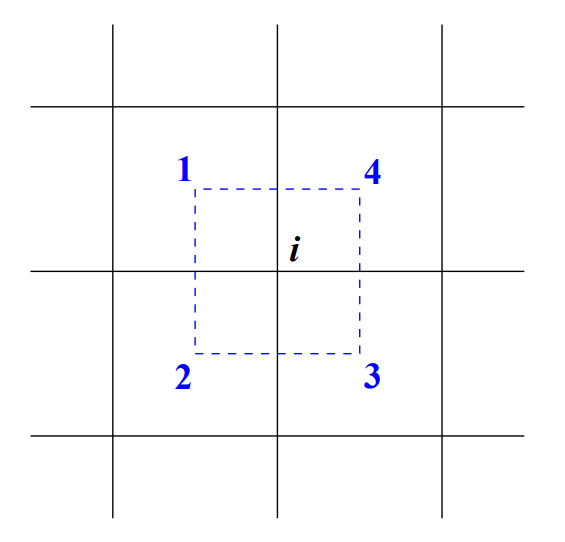}}
    \caption{(a) The 2D square lattice and its dual lattice. The black solid dots represent the 2D square lattice, while the white hollow dots represent its dual lattice. A pair of nearest-neighbor sites \(\langle i,j\rangle\) on the dual lattice intersects a bond of the original lattice, where \(r_{\mu}, \quad \mu=(i,\langle i,j\rangle)\) is defined. New \(\mathbf{Z}_2\) variables \(\sigma_i=\pm 1\) are defined on the dual lattice. (b) Four dual lattice sites corresponding to a single site of the original lattice.}
    \label{dual}
\end{figure}

In this section, we will solve for the phase transition temperature of the 2D Ising model through its self-duality and map the 2D classical statistics to the 1D quantum statistics using the transfer matrix, thereby deriving the quantum critical point of the 1D transverse Ising model. First, we write the Hamiltonian of the antiferromagnetic 2D Ising model as
\begin{equation}
H = -J \sum_{\langle i, j\rangle} S_i S_j, \quad J > 0
\end{equation}
where \(\langle i, j\rangle\) represents a pair of nearest-neighbor lattice sites \((i, j)\), and classical spins \(S_i = \pm 1\) are defined on site \(i\). Considering the square lattice, we can write its partition function and perform a series of transformations
\begin{equation}
Z = \sum_{\left\{S_j\right\}} \mathrm{e}^{K \sum_{\langle j, \ell\rangle} S_j S_{\ell}} = \sum_{\left\{S_j\right\}} \prod_{\langle j, \ell\rangle} \mathrm{e}^{K S_j S_{\ell}} = \sum_{\left\{S_j\right\}} \prod_{\langle j, \ell\rangle} \sum_{r=0}^1 C_r(K)\left(S_j S_{\ell}\right)^r, \quad K = \beta J
\end{equation}
where \(C_0(K) = \cosh K\) and \(C_1(K) = \sinh K\). Now, we need to introduce a \(\mathbf{Z}_2\) variable, or in other words, a variable \(r_{\mu}\) taking values \(\pm 1\), where \(\mu \equiv (i, \langle i,j\rangle)\). Consequently, we further rewrite the partition function in terms of \(r_{\mu}\)
\begin{equation}
\begin{aligned}
Z = \sum_{\left\{S_j\right\}} \sum_{\left\{r_\mu\right\}} \prod_{\langle j, \ell\rangle} C_{r_\mu}(K) \prod_i S_i^{\sum_{\langle i, j\rangle} r_\mu} \\
= \sum_{\left\{r_\mu\right\}} \prod_{\langle j, \ell\rangle} C_{r_\mu}(K) \prod_i \sum_{S_i = \pm 1} S_i^{\sum_{\langle i, j\rangle} r_\mu} \\
= \sum_{\left\{r_\mu\right\}} \prod_{\langle j, \ell\rangle} C_{r_\mu}(K) \prod_i 2 \delta\left[\bmod _2\left(\sum_{\langle i, j\rangle} r_\mu\right)\right]
\end{aligned}
\label{B3}
\end{equation}
where the Kronecker delta function \(\delta\) takes the value 1 when the variable is even and 0 when it is odd. By analyzing the above equations, we can see that the summation over \(r_{\mu}\) in the first equality actually includes the contributions from the four bonds connected to the site \(i\). The second equality transforms the summation over spin configurations into a summation over spins at each site. The third equality writes the partition function in terms of the variable \(r_{\mu}\) defined on the bonds. Now we define the dual lattice of the 2D square lattice as shown in \ref{dual}.
We define new \(\mathbf{Z}_2\) variables \(\sigma_i=\pm 1\) on the dual lattice. For each bond of the original lattice, there exists a pair of sites \(i,j\) on the dual lattice, on which we can define
\begin{equation}
r_\mu = \frac{1}{2}\left(1-\sigma_i \sigma_j\right)
\end{equation}
Next, we solve for the summation of \(r_{\mu}\) over the four nearest-neighbor bonds of a site in the original lattice, writing its general form as
\begin{equation}
\sum_{\langle i, j\rangle} r_\mu = 2 - \frac{1}{2}\left(\sigma_1 \sigma_2 + \sigma_2 \sigma_3 + \sigma_3 \sigma_4 + \sigma_4 \sigma_1\right)
\end{equation}
The 1, 2, 3, 4 dual lattice sites are shown in \ref{dual}(b). This summation has four possible cases:

i) \(\sigma_i = 1, \quad \forall i\), with two equivalent configurations that can be found by flipping all \(\sigma_i\)
\begin{equation}
\sum_{\langle i, j\rangle} r_\mu = 0
\end{equation}

ii) \(\sigma_1 = \sigma_2 = \sigma_3 = -\sigma_4 = 1\), with eight equivalent configurations
\begin{equation}
\sum_{\langle i, j\rangle} r_\mu = 2
\end{equation}

iii) \(\sigma_1 = \sigma_3 = -\sigma_2 = -\sigma_4 = 1\), with two equivalent configurations that can be found by flipping all \(\sigma_i\)
\begin{equation}
\sum_{\langle i, j\rangle} r_\mu = 4
\end{equation}

iv) \(\sigma_1 = \sigma_4 = -\sigma_2 = -\sigma_3 = 1\), with four equivalent configurations
\begin{equation}
\sum_{\langle i, j\rangle} r_\mu = 2
\end{equation}

Based on the four cases listed above, the delta function in \eqref{B3} always takes the value 1, so the partition function can be written as
\begin{equation}
Z = \frac{1}{2} 2^N \sum_{\left\{\sigma_i\right\}} \prod_{\langle j, \ell\rangle} C_{\left[\left(1-\sigma_i \sigma_j\right) / 2\right]}(K)
\label{B10}
\end{equation}
where \(\langle j, \ell\rangle\) is a pair of nearest-neighbor sites on the dual lattice, and the factor \(1/2\) comes from the fact that each \(r_{\mu}\) corresponds to two configurations. Now, we attempt to write the partition function in \eqref{B10} as a summation of Boltzmann factors. Here, we abbreviate \((1-\sigma_i \sigma_j)/2\) as \(r\)
\begin{equation}
\begin{aligned}
C_r(K) &= \cosh K[1 + r(\tanh K - 1)] \\
&= \cosh K \exp \{\log [1 + r(\tanh K - 1)]\} \\
&= \cosh K \exp (r \log \tanh K) \\
&= \cosh K \exp \left[\frac{1}{2}\left(1 - \sigma_i \sigma_j\right) \log \tanh K\right] \\
&= (\cosh K \sinh K)^{1/2} \exp \left(-\frac{1}{2} \log \tanh (K) (\sigma_i \sigma_j)\right)
\end{aligned}
\label{B11}
\end{equation}
Substituting \eqref{B11} into \eqref{B10} and noting that in the product over bonds, for a square lattice with \(N\) sites, there are \(2N\) bonds in total. We can obtain the form of the partition function of the dual lattice as
\begin{equation}
Z = \frac{1}{2}(\sinh 2 \tilde{K})^{-N} \sum_{\left\{\sigma_i\right\}} \exp \left(\tilde{K} \sum_{\langle j, \ell\rangle} \sigma_j \sigma_{\ell}\right)
\label{B12}
\end{equation}
where \(\tilde{K} \equiv -\frac{1}{2} \log \tanh K\). From \eqref{B12}, we can see that the dual of the 2D Ising model gives a partition function of the same form as the original model, thus it is called self-dual. Dual transformations are often used to study the high-temperature region (here corresponding to relatively small \(K\)) properties of a model to reflect the properties of its dual model in the low-temperature region (here corresponding to relatively large \(K\)). This is because the high-temperature region can usually be more conveniently studied through techniques like high-temperature expansion, while the properties of the low-temperature region are relatively more difficult to obtain (for example, in Monte Carlo simulations, the low-temperature region often encounters issues with local minima). Here, \(K \rightarrow 0\) corresponds to \(\tilde{K} \rightarrow +\infty\), and vice versa. To find the critical point (or the phase transition temperature) of the system, we need to find the singularity of the free energy of the system, averaged over each lattice site, which is
\begin{equation}
f = -\frac{1}{N} \log Z
\end{equation}
From (B.12), we have
\begin{equation}
f(K) = \log \sinh 2 \tilde{K} + f(\tilde{K})
\end{equation}
Since \(\sinh K\) is an analytical function, and \(\tilde{K}(K)\) is a monotonous function, the singularity of \(f(K)\) must be the singularity of \(f(\tilde{K})\), i.e., \(K_c = \tilde{K}_c\). Thus, we can solve for the critical point of the 2D Ising model
\begin{equation}
\mathrm{e}^{2 K_c} = \frac{\mathrm{e}^{2 K_c} + 1}{\mathrm{e}^{2 K_c} - 1} \Rightarrow K_c = \frac{1}{2} \log (1 + \sqrt{2})
\label{B15}
\end{equation}
The above derivation is for the case where the coupling in the \(x\) and \(y\) directions of the 2D Ising model is the same, i.e., \(K_x = K_y\). For the case of \(K_x \neq K_y\), it can be found that the \(x\)-direction coupling of the dual lattice corresponds to the \(y\)-direction coupling of the original lattice, and vice versa. Therefore, through the coupling relationship in the isotropic condition \(\tilde{K} = -\frac{1}{2}\log(\tanh K)\), it can be generalized to the anisotropic case
\begin{equation}
\tilde{K}_y \equiv -\frac{1}{2} \log \tanh K_x, \quad \tilde{K}_x \equiv -\frac{1}{2} \log \tanh K_y
\end{equation}
Thus, the critical point derived from \eqref{B15} transforms into a boundary of the phase transition from the paramagnetic phase to the ferromagnetic phase, expressed as
\begin{equation}
\sinh \left(2 K_{x c}\right) \sinh \left(2 K_{y c}\right) = 1
\label{B17}
\end{equation}
Next, we will use the transfer matrix method to map the anisotropic 2D Ising model to the 1D transverse Ising model. We still start from the form of the partition function
\begin{equation}
Z = \sum_{\left\{S_i\right\}} \mathrm{e}^{-S}, \quad S = -K \sum_{\langle i, j\rangle} S_i S_j
\end{equation}
Introduce coordinates \((p,q)\) representing the positions in the \(x\) and \(y\) directions, respectively. Suppose there are \(N\) sites in the \(x\) direction and \(M\) sites in the \(y\) direction, and decompose the action \(S\) as
\begin{equation}
\begin{gathered}
S = \sum_{q=1}^M L(q, q+1) \\
L(q, q+1) = \sum_{p=1}^N\left(-K_x S_{p, q} S_{p+1, q} - K_y S_{p, q} S_{p, q+1}\right)
\end{gathered}
\end{equation}
First, we consider the terms related to \(K_y\). At this point, we can ignore \(K_x\) and temporarily set it to 0
\begin{equation}
Z_p = \sum_{S_{p, 1}, \ldots, S_{p, M}} \prod_q T_{p q}^y, \quad T_{p q}^y = \mathrm{e}^{K_y S_{p, q} S_{p, q+1}}
\end{equation}
To express the four possible configurations, i.e., \(S_{p,q}=\pm1, S_{p,q+1}=\pm1\), we introduce the eigenvectors of \(\sigma^z\) as basis vectors, i.e.,
\begin{equation}
S_{p, q} = +1 \rightarrow \binom{1}{0}, \quad S_{p, q} = -1 \rightarrow \binom{0}{1}
\end{equation}
Then the transfer matrix can be written as
\begin{equation}
T_{p,q}^y = \left(\begin{array}{ll}
\mathrm{e}^{K_y} & \mathrm{e}^{-K_y} \\
\mathrm{e}^{-K_y} & \mathrm{e}^{K_y}
\end{array}\right) = \mathrm{e}^{K_y} \mathbf{1} + \mathrm{e}^{-K_y} \sigma_{p q}^x = \mathrm{e}^{K_y}\left(\mathbf{1} + \mathrm{e}^{-2 K_y} \sigma_{p q}^x\right)
\end{equation}
To write the transfer matrix in the form of a matrix exponential, we use the exponential property of \(\sigma^x\) and the dual relationship of the coupling constant
\begin{equation}
\begin{aligned}
\exp \left(\tilde{K}_y \sigma^x\right) &= \cosh \tilde{K}_y + \sinh \tilde{K}_y \sigma^x = \cosh \tilde{K}_y \left(1 + \tanh \tilde{K}_y \sigma^x\right) \\
\tanh \tilde{K}_y &= \mathrm{e}^{-2 K_y}
\end{aligned}
\label{B23}
\end{equation}
Thus, the transfer matrix can be written as
\begin{equation}
T_{p q}^y = \left(\sinh \tilde{K}_y \cosh \tilde{K}_y\right)^{-1 / 2} \exp \left(\tilde{K}_y \sigma_{p q}^x\right) = \left(2 \sinh 2 K_y\right)^{1 / 2} \exp \left(\tilde{K}_y \sigma_{p q}^x\right)
\end{equation}
Next, we solve for the transfer matrix \(T_{pq}^x\) related to \(K_x\). Using ket and bra vectors to represent the involved spins, we have
\begin{equation}
\left\langle S_{p, q} S_{p+1, q}\left|T_{p q}^x\right| S_{p, q+1} S_{p+1, q+1}\right\rangle = \mathrm{e}^{K_x S_{p, q} S_{p+1, q}}
\end{equation}
Thus, we have
\begin{equation}
\begin{aligned}
& T_{p q}^x\left|S_{p, q} = 1, S_{p+1, q} = 1\right\rangle = T_{p q}^x\left|S_{p, q} = -1, S_{p+1, q} = -1\right\rangle = \mathrm{e}^{K_x} \\
& T_{p q}^x\left|S_{p, q} = 1, S_{p+1, q} = -1\right\rangle = T_{p q}^x\left|S_{p, q} = -1, S_{p+1, q} = 1\right\rangle = \mathrm{e}^{-K_x}
\end{aligned}
\end{equation}
It can be found that \(T_{pq}^x\) does not contain any information about the spin state at \(q+1\). Combining the forms of the transfer matrices in the \(x\) and \(y\) directions, we can summarize the partition function of the 2D Ising model as follows
\begin{equation}
\begin{gathered}
Z = \left(2 \sinh 2 K_y\right)^{N M / 2} \operatorname{Tr} T^M \\
T = \exp \left(K_x \sum_p \sigma_p^z \sigma_{p+1}^z\right) \exp \left(\tilde{K}_y \sum_p \sigma_p^x\right)
\end{gathered}
\label{B27}
\end{equation}
With the complete transfer matrix, we can obtain an exactly solvable fermion model through the Jordan-Wigner transformation. However, here we will use another method—treating the transfer matrix as the evolution operator in quantum mechanics \cite{PhysRevD.17.2637}. In the transfer matrix in \eqref{B27}, multiply each variable in the exponent by a small imaginary time interval \(\Delta \tau\) and expand it to the first order, then rewrite it in the form of an exponent to eventually obtain an equivalent Hamiltonian. The error of this process is of the order \(\mathcal{O}\left(\Delta \tau^2\right)\), which is almost negligible when \(\Delta \tau \rightarrow 0\). Thus, we need
\begin{equation}
K_x \propto \Delta \tau, \quad \mathrm{e}^{-2 K_y} \propto \Delta \tau
\end{equation}
In other words, in this way, we focus on the region where \(K_x \ll 1\) and \(K_y \gg 1\). In this region, the system has two phases: the paramagnetic and ferromagnetic phases. Now treat the finite temperature \cite{10.1143/PTP.56.1454} as the result of imaginary time evolution
\begin{equation}
\tilde{K}_y = \Delta \tau, \quad K_x = \lambda \Delta \tau
\end{equation}
where \(\lambda\) is an arbitrarily chosen constant. Substituting these settings into \eqref{B27}, we can obtain the partition function in the form
\begin{equation}
Z \propto \operatorname{Tr} \mathrm{e}^{-\beta_{Q M} H}, \quad \beta_{Q M} = M \Delta \tau
\end{equation}
where the equivalent Hamiltonian \(H\) is
\begin{equation}
H = -\sum_p \sigma_p^x - \lambda \sum_p \sigma_p^z \sigma_{p+1}^z
\end{equation}
Considering the thermodynamic limit, i.e., \(M, N \rightarrow \infty\), it can be found that \(\beta_{QM} \rightarrow \infty\), meaning that a quantum phase transition (which is independent of temperature but related to the coupling constants in the Hamiltonian) exists from the magnetically ordered to the disordered state only when the equivalent temperature of the system tends to zero (the system is in the ground state). In the region we discuss, \(K_x, K_y\), \eqref{B17} can be expanded as
\begin{equation}
2 K_{x c} \frac{1}{2} \mathrm{e}^{2 K_{y c}} = 1
\label{B32}
\end{equation}
Combining this with the second equation of \eqref{B23}, we obtain
\begin{equation}
K_x = \lambda \mathrm{e}^{-2 K_y}
\end{equation}
Combining this equation with \eqref{B32}, we find that the quantum critical point is \(\lambda_c = 1\). When \(\lambda > \lambda_c\), the ground state of the system is magnetically ordered, dominated by the antiferromagnetic Ising interaction; when \(\lambda < \lambda_c\), the ground state of the system is disordered, mainly due to the large transverse field term giving significant quantum fluctuations of the ground state magnetization.
\section{}
\subsection{A Brief Review of Phase Transitions and Scaling Laws}
\label{A Brief Review of Phase Transitions and Scaling Laws}
Scaling laws refer to the invariance of certain properties of a physical system under changes in scale (for lattice systems, the size of the lattice). In physics, especially in critical phenomena, scaling laws manifest as various physical quantities, such as correlation length, magnetization, and specific heat, exhibiting power-law behavior as system parameters (e.g., temperature, magnetic field) change near the critical point. These critical exponents are influenced by some universal properties, rather than microscopic details. For spin systems, these properties might include system size, interaction range, and spin dimensionality. Common scaling relations can be seen in \ref{tabC1} and \ref{tabC2} \cite{enwiki:1218969478}.

\begin{table}[H]
\centering
\caption{Physical Quantities in Critical Phenomena and Their Symbolic Representations}
\label{tabC1}
\begin{tabular}{|c|l|}
\hline
\textbf{Symbol} & \textbf{Description} \\
\hline
$\Psi$ & order parameter \\
\hline
$\tau$ & reduced temperature minus 1, $\frac{T-T_{\mathrm{c}}}{T_{\mathrm{c}}}$ \\
\hline
$f$ & specific free energy \\
\hline
$C$ & specific heat; $-T \frac{\partial^2 f}{\partial T^2}$ \\
\hline
$J$ & source field \\
\hline
$\chi$ & the susceptibility, compressibility, etc.; $\frac{\partial \psi}{\partial J}$ \\
\hline
$\xi$ & correlation length \\
\hline
$d$ & the number of spatial dimensions \\
\hline
$\langle\psi (\vec{x}) \psi (\vec{y})\rangle$ & the correlation function \\
\hline
$r$ & spatial distance \\
\hline
\end{tabular}
\end{table}

\begin{table}[H]
\centering
\caption{Power-law Behavior of Physical Quantities Near the Critical Point}
\label{tabC2}
\begin{tabular}{|c|c|c|c|c|c|}
\hline \multicolumn{2}{|c|}{\begin{tabular}{c} 
Critical exponents for \\
$\tau>0$ (disordered phase)
\end{tabular}} & \multicolumn{2}{|c|}{\begin{tabular}{l} 
Critical exponents for \\
$\tau<0$ (ordered phase)
\end{tabular}}& \multicolumn{2}{|c|}{\begin{tabular}{l} 
Critical exponents for \\
$\tau<0$ (ordered phase)
\end{tabular}} \\
\hline Greek letter & relation & Greek letter & relation &Greek letter & relation\\
\hline$\alpha$ & $C \propto \tau^{-\alpha}$ & $\alpha^{\prime}$ & $C \propto(-\tau)^{-\alpha^{\prime}}$ &$\delta$ &$J \propto \Psi^\delta$ \\
\hline$\gamma$ & $\chi \propto \tau^{-\gamma}$ & $\beta$ & $\Psi \propto(-\tau)^\beta$ & $\eta$ &$\langle\psi(0) \psi(r)\rangle \propto r^{-d+2-\eta}$ \\
\hline$v$ & $\xi \propto \tau^{-v}$ & $\gamma^{\prime}$ & $\chi \propto(-\tau)^{-\gamma^{\prime}}$ & & \\
\hline & & $v^{\prime}$ & $\xi \propto(-\tau)^{-v^{\prime}}$ & & \\
\hline
\end{tabular}
\end{table}

This universal behavior of systems leads to the concept of universality classes. Systems with different microscopic structures and interactions exhibit the same critical behavior if they have the same symmetry and spatial dimension at the critical point. These systems display the same scaling behavior and critical exponents macroscopically.

Analysis of some experimental data has revealed certain constraint relations among these critical exponents, known as scaling relations. For any continuous phase transition system, the following three scaling relations hold:
\begin{equation}
    \begin{aligned}
& \alpha+2 \beta+\gamma=2 \\
&  \gamma=\beta(\delta-1) \\
& (2-\eta) \nu=\gamma \\
\end{aligned}
\end{equation}
These three scaling laws are known as the Rushbrooke scaling law \cite{10.1063/1.1734338}, the Griffiths scaling law, and the Fisher scaling law \cite{PhysRevLett.56.416}, respectively. For systems with spatial dimensions less than the upper critical dimension, there exists a hyperscaling relation, also known as the Josephson scaling law
\begin{equation}
    2-\alpha=\nu d
\end{equation}
For some common models or theories, the critical exponents can be found in the table below \cite{lzu_phy}.

\begin{table}[H]
\centering
\caption{Critical Exponents for Some Typical Theories or Models}
\label{tab:critical_phenomena}
\begin{tabular}{|c|c|c|c|c|c|c|}
\hline Type & $\alpha$ & $\beta$ & $\gamma$ & $\delta$ & $\nu$ & $\eta$ \\
\hline Landau Mean Field & 0 & $\frac{1}{2}$ & 1 & 3 & $\frac{1}{2}$ & 0 \\
Gaussian Theory & $2-\frac{d}{2}$ & $\frac{d-2}{4}$ & 1 & $\frac{d+2}{d-2}$ & $\frac{1}{2}$ & 0 \\
2D Ising Model & 0 & $\frac{1}{8}$ & $\frac{7}{4}$ & 15 & 1 & $\frac{1}{4}$ \\
3D Ising Model & $0.110(1)$ & $0.3265(3)$ & $1.2372(5)$ & $4.789(2)$ & $0.6301(4)$ & $0.0364(5)$ \\
\hline
\end{tabular}
\end{table}
The critical exponents of the 3D Ising model have not been analytically solved yet, and these values are obtained through Monte Carlo methods. The critical exponents of the 3D Ising model can also be obtained through the operator product expansion (OPE) in conformal field theory (CFT) \cite{Kos_2016}.

\end{document}